\documentclass[9pt,twocolumn,twoside]{pnas-new}
\templatetype{pnasresearcharticle} 
\title{Elastic avalanches reveal marginal behaviour in amorphous solids}
\author[a,b]{Baoshuang Shang}
\author[a,1]{Pengfei Guan} 
\author[b,1]{Jean-Louis Barrat}
\affil[a]{Beijing Computational Science Research Center, Beijing 100193,  China}
\affil[b]{Univ. Grenoble Alpes, CNRS, LIPhy, 38000 Grenoble, France}
\leadauthor{Shang}
\significancestatement{At large strains,  glassy systems deform through a series of elastic loading phases followed by energy and stress drops without a characteristic scale, similar to earthquakes. Less is known on the small deformation case, which is related to the structure of the energy landscape close to an energy minimum. We investigate numerically this regime, and show that even at very small strains the deformation proceeds through avalanches that are power-law distributed, with a universal exponent that corresponds to the predictions of mean field theory for a hierarchical phase space structure. These avalanches reveal marginal stability in the amorphous solid, which is intrinsically inelastic. By investigating the preparation and size dependence, we infer that the effect persists in the thermodynamic limit.
}
\authorcontributions{B.S.S,P.F.G and J.L.B designed the study, B.S.S performed the research, B.S.S,P.F.G and J.L.B analyzed the results, B.S.S and J.L.B wrote the paper.}
\authordeclaration{
The authors declare no conflict of interest.}
\correspondingauthor{\textsuperscript{1}To whom correspondence should be addressed. E-mail:  pguan@csrc.ac.cn \& jean-louis.barrat@univ-grenoble-alpes.fr }


\begin{abstract}
  Mechanical deformation of amorphous solids can be described as consisting of an
  elastic part in which the stress increases linearly with strain, up to a yield point at which the solid either fractures or starts deforming plastically. It is well established, however, that the apparent linearity of stress with strain is actually a proxy for a much more complex behavior, with a microscopic plasticity that is reflected in diverging nonlinear elastic coefficients. Very generally, the complex structure of the energy landscape is expected to induce a singular response to small perturbations. In the athermal quasistatic regime,  this response manifests itself in the form of a scale free plastic activity. The distribution of the corresponding avalanches should reflect, according to theoretical mean field calculations (Franz and Spigler, Phys. Rev. E., 2017, 95, 022139), the geometry of phase space in the vicinity of a typical local minimum. In this work, we characterize this distribution for simple models of glass forming systems, and we find that its scaling is compatible with the mean field predictions for systems above the jamming transition. These systems exhibit marginal stability, and scaling relations that hold in the stationary state are  examined and confirmed in the elastic regime. By studying the respective influence of system size and age,  we suggest that marginal stability is systematic in the thermodynamic limit.
\end{abstract}

\dates{This manuscript was compiled on \today}
\doi{\url{www.pnas.org/cgi/doi/10.1073/pnas.XXXXXXXXXX}}

\begin{document}

\maketitle
\thispagestyle{firststyle}
\ifthenelse{\boolean{shortarticle}}{\ifthenelse{\boolean{singlecolumn}}{\abscontentformatted}{\abscontent}}{}

\dropcap{T}he response of amorphous solids and yield stress fluids  to a mechanical deformation has attracted a considerable attention from the statistical physics as well as materials science community in the recent years. A large number of numerical and theoretical studies  have been devoted to the regime of stationary plastic flow, and particularly in the limit of zero strain rate and negligible thermal effects, the so-called athermal quasi static (AQS) regime. In this regime, it is now well accepted that the flow proceeds by local instabilities called shear transformations, that interact elastically and can organise in larger scale events called avalanches. Each event results, at constant strain, into a stress  or energy drop.  The statistics of these drops are typically a power law with a cutoff that depends on system size. This behavior can be described in terms of simple elasto-plastic models, in which subvolumes of the glass are described as  linear  elastic elements that yield above some critical stress, possibly triggering the yield of other elements as the stress is transmitted through the system by an elastic propagator. This simplified picture, while very successful in describing the collective behavior at large deformation, completely ignores the fine structure of the energy landscape. This structure is effectively responsible for the dynamics of the local yield process, which in these models is described in terms of some effective damping parameter.

Another set of studies has focused on the yield process itself, i.e. the transition from an essentially reversible deformation towards irreversible  plastic flow or failure. This transition has been shown to depend critically on the thermal history of the system, with poorly annealed systems undergoing a rather smooth transition to a flowing state, while very well annealed systems fracture abruptly \cite{Ozawa2018random}. This difference however is not directly related to the structure of phase space in the vicinity of a given minimum, as it only occurs at large deformations.

Finally,  considerable attention has been devoted recently to the possible existence of so-called ``marginally stable'' glassy states, in which the local phase space has a hierarchical organisation that can be associated, in the language of spin glasses, with a full breaking of the replica symmetry. Ordinary glasses, on the other hand, have a simpler energy landscape, with many minima separated by rather large barriers.
{The transition towards marginally stable glasses was first explored in hard sphere systems \cite{berthier2016growing,Charbonneau2017,PhysRevX.9.011049}, and some signatures of this structure have been observed in recent simulations of soft spheres \cite{Scalliet2019,PhysRevLett.122.255502}, however with a limitation to finite range interactions.}

In this context, it was shown  by Spigler and Franz \cite{PhysRevE.95.022139} that the hierarchical  structure of phase space should result in a peculiar response to mechanical deformation,  somewhat similar to the one observed in flowing systems. Shear transformations associated with stress or energy drops  have long been observed  in the elastic regime at low temperature both in experiment and simulation \cite{Papakon2008, PhysRevLett.112.155501,denisov2017universal,lagogianni2018plastic}. However, the  events observed at small deformation are generally localized, partly reversible and thermal-history dependent, in contrast with the steady state case \cite{Jin2018stability,PhysRevLett.93.016001,PhysRevE.82.055103,PhysRevE.95.022611}. Organisation in avalanches displaying a power law distribution is observed only at  large strains close to the yield point, the exponent of the corresponding power law being still controversial  \cite{lagogianni2018plastic,krisponeit2014crossover,leishangthem2017yielding,regev2015reversibility}.  In contrast, the prediction made in ref. \cite{PhysRevE.95.022139} is that even at very small strains (vanishingly small in the mean field calculation), the events are scale free avalanches with a distribution that reflects the structure of phase space described by the Parisi function. A very specific prediction is made concerning the exponent of the distribution of avalanche sizes in the mean field limit, and preliminary numerical \cite{PhysRevE.95.022139} {and experimental \cite{peng2019anomalous}} results were shown to be close to this prediction.
In this work, we investigate avalanche statistics far below yielding,
both in two and three dimensions, for a system of particles interacting through a  Lennard-Jones potential.  Different system sizes and  different thermal histories of the initial configuration are considered. The simulations are carried out using the  athermal quasistatic protocol (AQS) in simple shear, volume conserving deformations.
We find that even at very  small strains, the mean value of avalanche size is sub-extensive with system size, with a  finite size scaling exponent that depends on thermal-history.
By making a simple scaling ansatz, all the data for the avalanche size distribution can be collapsed onto a single master curve, with  a universal avalanche exponent in the transient state clearly distinct from the one observed previously in the steady  plastic flow regime,  and close to the value predicted in ref. \cite{PhysRevE.95.022139}.  The mean field calculation, being done in the limit of infinite dimensions, does not convey any information concerning the spatial structure and physical nature of the corresponding events in real space. We therefore investigate the possibility that the  marginally stable state is amenable to an elasto-plastic description \cite{PhysRevX.6.011005} involving  interacting zones characterized by a ``pseudo-gap'' . Within this framework,
a universal scaling relation observed in the steady state is also valid in the transient state, and directly connects the avalanche energy with the dissipation in the transient state at zero temperature and with the exponents characterizing the pseudo gap associated with marginal stability. The latter is also  found to behave as predicted by the elasto-plastic  models studied in \cite{PhysRevX.6.011005} .
By analysing the dependence of the results on thermal history and size, we infer that, in the thermodynamic limit, the amorphous solid shows intrinsic inelastic behavior.
\section*{Results}
\subsection*{Avalanche  number density }
To investigate the statistics of avalanches, we use the avalanche number  density $R(S,N,T_\text{ini})$ \cite{PhysRevLett.109.105703}, which is defined as the number of avalanches per unit of avalanche size (here measured by the corresponding energy drop)  $S$ and per unit strain. $N$ and $T_\text{ini}$ refer to the system size (number of atoms) and the initial temperature from which the system has been quenched to zero temperature.
{For reference, the mode coupling temperature is} $0.325$ in 2D and  $0.435$ in 3D system.
The corresponding normalized avalanche distribution $P(S)$ and total avalanche number $M$ per unit strain  are 
\begin{equation}
  M(N,T_\text{ini}) = \int_0^\infty R(S) dS   \ \ ; P(S,N,T_\text{ini}) = R(S)/M
\end{equation}
and the total energy  per unit strain dissipated in avalanches is 
\begin{equation}
  \eta(N,T_\text{ini}) = \int_0^\infty S R(S,N,T_\text{ini}) dS = M\langle  S  \rangle . 
\end{equation} 

The systems and procedure are described in the Methods section.
Here we only recall that the range of strains used to collect the statistics is $\gamma \in [0,0.02]$, much below the yield strain $\gamma_\text{Y}$ ($\gamma_\text{Y} \approx 0.06$ and $0.08$ in 2D,3D systems, respectively). In this regime, we have checked that 
$R(S,N,T_\text{ini})$ is insensitive to the strain interval used to collect the statistics, as illustrated in Fig. \ref{fig:d} in Supplamentary Information(SI). 

\begin{figure}[!htbp]
  \centering
  \includegraphics[width=1.0\columnwidth]{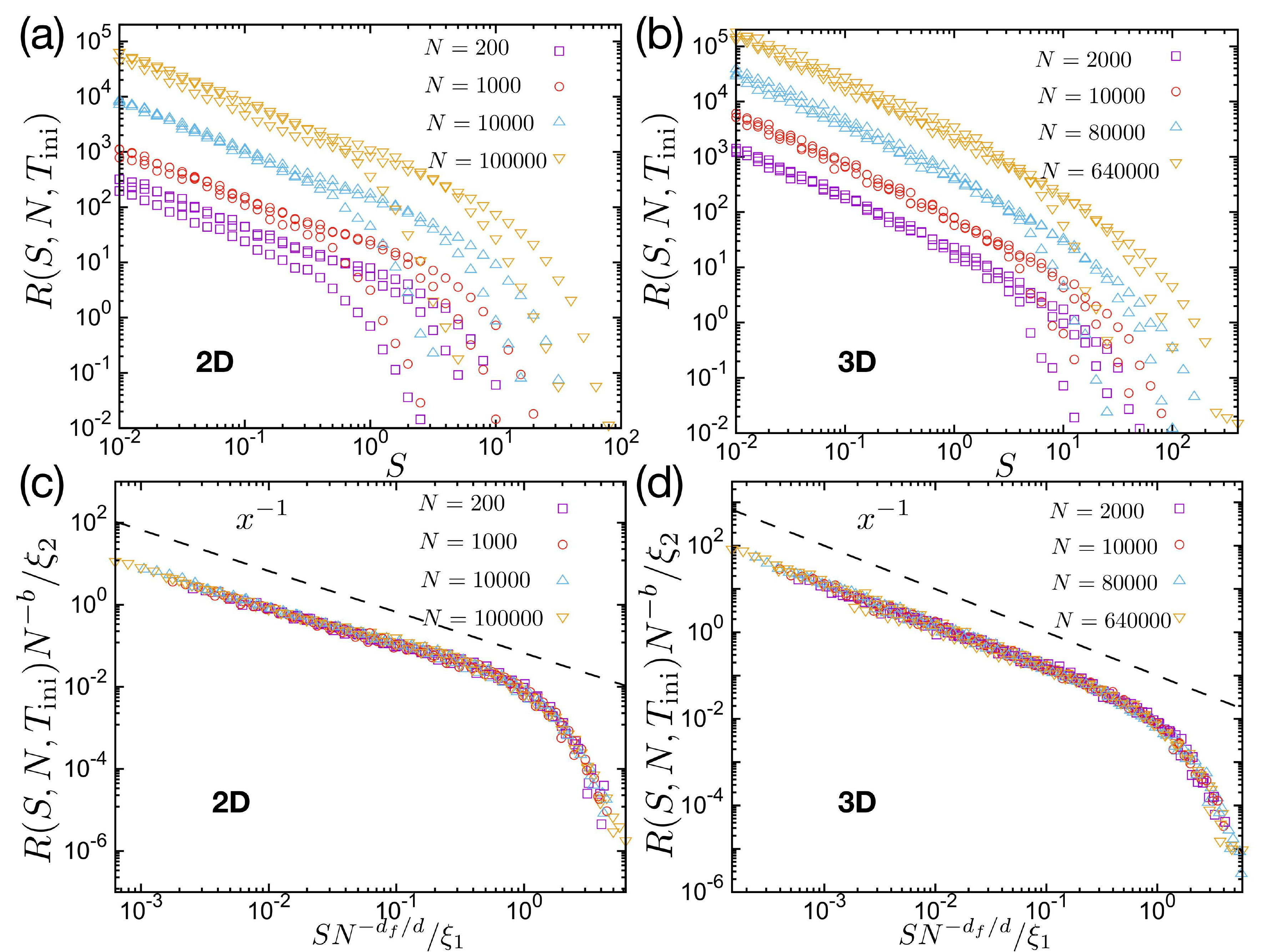}
  \caption{\textbf{Avalanche number density versus avalanche size} (a),(b): Avalanche  number density  for  different system sizes and thermal histories in 2D and 3D, respectively. (c),(d): data collapse using two exponents,  ${{d_f}/{d}}$ and $b$ for the scaling as a function of system size, and two prefactors $\xi_1$,$\xi_2$ that depend on thermal history (see text). The  parameters are  fitted from figure \ref{fig:2}.
  The dashed line shows the avalanche exponent $-1$ predicted by  mean field theory\cite{PhysRevE.95.022139} near the ground state.
  The investigated strain range is far below yield strain, $\gamma \in [0,0.02]$, where the  yield strain $\gamma_\text{Y} \approx 0.06$ and $0.08$ in 2D and 3D, respectively.}
  \label{fig:1}
\end{figure}

The number density  of avalanches is shown  as a function of their size in Figure \ref{fig:1}. It displays  a typical power-law distribution with a cutoff that depends on system size and on thermal history. 
In contrast with the case of stationary plastic flow \cite{PhysRevLett.93.016001}, avalanches in the transient state are influenced by thermal history, which also determines   the brittleness of the amorphous material \cite{PhysRevLett.95.095502,Ozawa2018random}.
From figure \ref{fig:1} and Fig. \ref{fig:c}
, it is seen that, for a given system size, the cutoff value and the extent of the  power law behavior in the  $R$ become smaller as $T_\text{ini}$ decreases and the stability of the initial configuration increases.

\subsection*{Scaling analysis}
Generally, one expects that the avalanche number density $R(S,N,T_\text{ini})$ can be described as a power law distribution with cutoff caused by finite size effects, i.e. $ R(S,N,T_\text{ini}) \sim S^{-\tau} f(S/S_c)$, where $S_c$ is the cutoff value influenced by system size and thermal history,  $\tau$ is the  avalanche exponent, and $f(S/S_c)$ is a cutoff function.  
Introducing the reduced size $\chi= S/S_c$, we can make the following scaling hypothesis:
\begin{equation}
  S_c  \sim \xi_1  N^{{d_f}/{d}} 
\end{equation}
\begin{equation}
  R(S,N,T_\text{ini}) \sim \xi_{2} N^{b} \chi^{-\tau}f(\chi)
\end{equation}

\begin{figure}
  \centering
  \includegraphics[width=1.0\columnwidth]{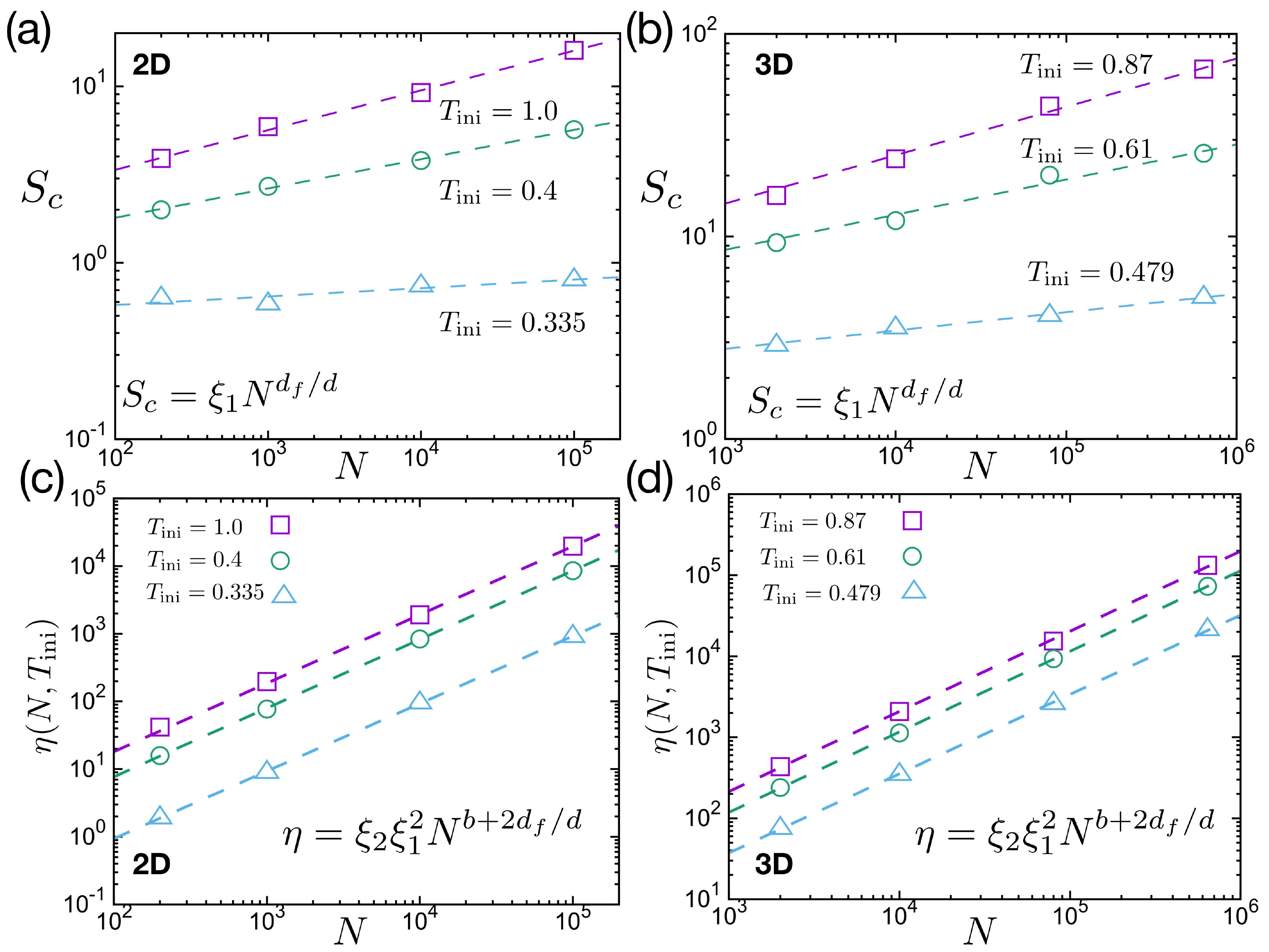}
  \caption{\textbf{Cutoff of the distribution of avalanche sizes and total avalanche energy.}(a),(b) Cutoff value $S_c$ versus system size in 2 and 3 dimensions  and for different thermal histories, the dashed line is a fit to a power law   $S_c=\xi_1 N^{d_f/d}$
  (c),(d) total avalanche energy versus system size, the dashed line is a  power law $\eta(N,T_\text{ini})=\xi_1^{2}\xi_2 N^{b+2d_f/d}$ where $d_f/d,\xi_1$  are obtained from fitting $S_c$.}
  \label{fig:2}
\end{figure}

Here $\xi_{1}$,$\xi_{2}$ are  prefactors which are determined by  thermal history, the exponents $d_f/d$ and $b$, which a priori could also depend on thermal history, describe the   cutoff due to the  finite size of the system. Here  $d_f$ is the fractal dimension of avalanches and $d$ is the dimension of system.  
The total avalanche energy per unit strain $\eta(N,T_\text{ini})$ can then be written as follows:
\begin{equation}
  \eta(N,T_\text{ini})  \equiv  \int_0^\infty R(S,N,T_\text{ini}) S d S 
  \sim \xi_{1}^2 \xi_{2} N^{b+2{d_f}/{d}} 
  \label{eq:energy}
\end{equation}
For values of the  avalanche exponent $\tau <2$, the cutoff value $S_c$ can be obtained from: 
\begin{equation}
  S_c = \frac{\int_0^\infty R(S,N,T_\text{ini}) S^2 dS}{\int_0^\infty R(S,N,T_\text{ini}) S d S} \sim  \xi_{1} N^{{d_f}/{d}} 
\end{equation}
One can then collapse the data onto a master curve, removing  the dependence on thermal history and system size. The parameters associated with system size ($d_f/d$ and $b$) and with thermal history ($\xi_1$ and $\xi_2$), can be fitted by the formula: $S_c=\xi_1 N^{d_f/d}$ and $\eta(N,T_\text{ini})=\xi_1^2 \xi_2 N^{b+2d_f/d}$, both for 2D and 3D systems, as shown in Figure \ref{fig:2}.  The values of the fit parameters are given in 
tables S1 and S2 in SI Appendix.
In figure \ref{fig:1}(c),(d), after data collapse, the avalanche number density shows a universal behavior and the avalanche exponent is close to unity  for 2D ($\tau=0.98 \pm 0.01$) and 3D ($\tau=1.01 \pm 0.01$) systems.  The fitting curve is shown in 
Fig. \ref{fig:e}.
\begin{figure}[!htbp]
  \centering
  \includegraphics[width=1.0\columnwidth]{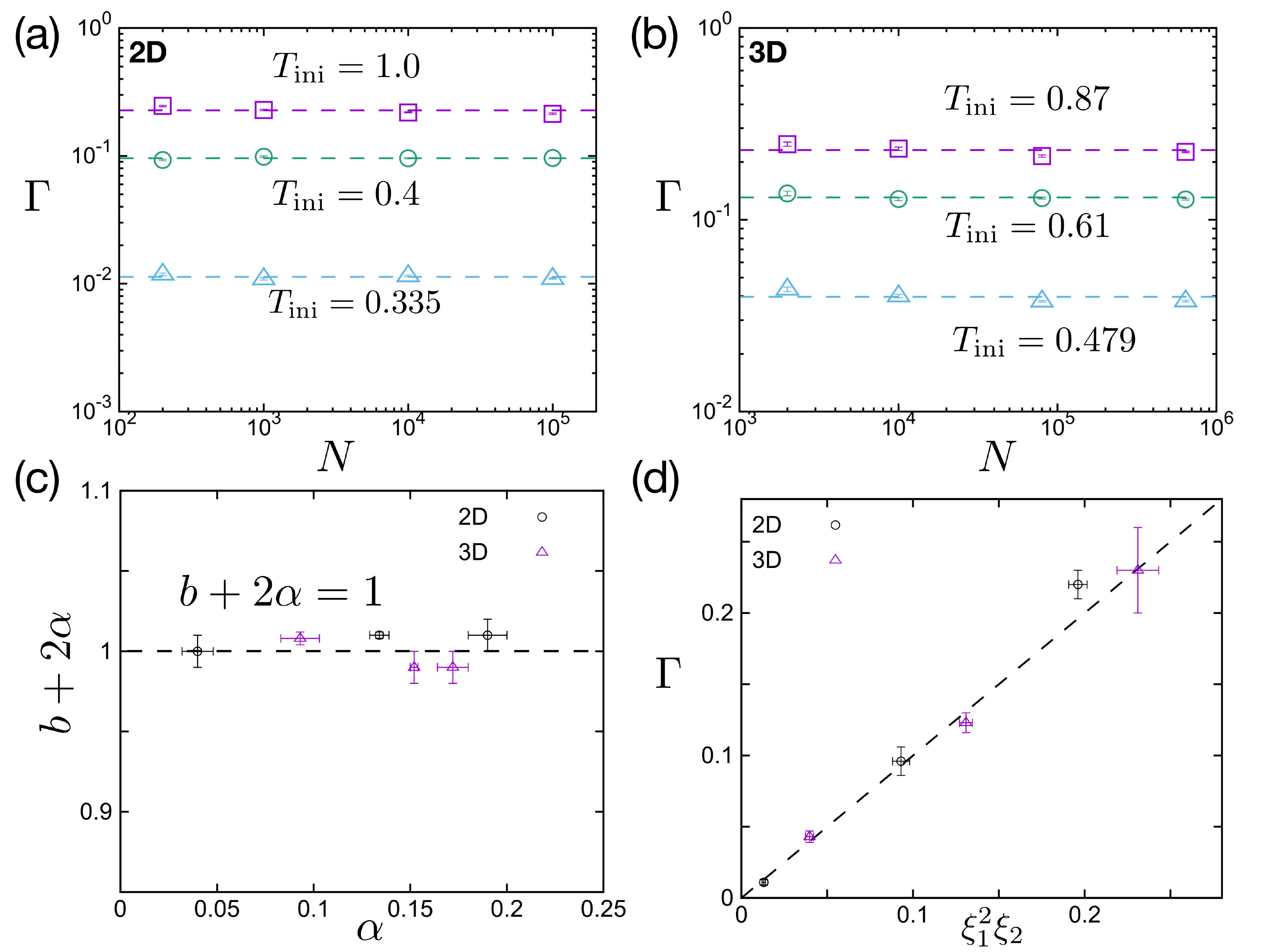}
  \caption{\textbf{Dissipated energy.}(a),(b) density of dissipated energy  versus system size in 2D and 3D, respectively. The horizontal dashed line is the mean value of $\Gamma$ for different sizes (c) universal finite size scaling relation.(d) Correlation between  energy dissipation  and avalanche energy densities. The dashed line is $y=x$. For $\Gamma$, the strain range is $\gamma \in [0,0.02]$.}
  \label{fig:3}
\end{figure} 
\subsection*{Energy balance and relation between exponents}
In order to check the consistency of the scaling analysis, it is worthwhile to consider it in relation with energy balance arguments. The total  energy  per unit strain of the avalanches is $\eta(N,T_\text{ini})$ (equation \ref{eq:energy}), which depends on  system size and on thermal history. As the system is deformed at zero temperature without thermostat, the avalanches constitute  the only mechanism that  dissipates energy, therefore energy balance implies that adding up this dissipated energy to the work done on the system  during loading should give the difference in energy between an initial state at zero strain and the final strain after straining by an amount $\gamma$. In other words, if $N\Gamma(\gamma)$ is the total energy dissipated in the process (which we expect to be extensive) one has the identity: 
\begin{equation}
  N\gamma^{-1}(U(\gamma)- U(0)) =N\gamma^{-1}\rho^{-1} {\int^{\gamma}_{0} {\sigma(\gamma)} d \gamma} - N \Gamma(\gamma)
  \label{eq:dissip}
\end{equation}
$\Gamma(\gamma)$ is the density of dissipated energy. It can be calculated from the stress strain curve using equation \ref{eq:dissip}. The data displayed in figure \ref{fig:3}(a) and \ref{fig:3}(b), show that this quantity is indeed independent of system size, as expected. If we now identify $N\Gamma$ with the total energy of the avalanches,  we obtain the universal scaling relation:
\begin{equation}
  \eta(N,T_\text{ini}) \sim N^{b+2d_f/d} \sim N
\end{equation}
and the relation between exponents:  $b+2d_f/d =1 $. Figure \ref{fig:3}(c) shows that this relationship,  which was first obtained in the plastic flow regime by Salerno \textit{et al} \cite{PhysRevLett.109.105703,PhysRevE.88.062206},  also holds for the transient avalanches in the elastic regime. Note that the  expected relation  for the prefactors, 
$\Gamma=\xi_{1}^2\xi_{2}$ is also confirmed in figure \ref{fig:3}(d).

\subsection*{Conclusion concerning the avalanche exponent} 
The avalanche exponent in our system is consistent with the theoretical work of Franz and Spigler \cite{PhysRevE.95.022139}, who confirmed their prediction by preliminary simulations of soft elastic spheres above jamming. Our system of Lennard-Jones particles with attractions  is significantly different, so that the result suggests a universal exponent for avalanches in the elastic regime,  independent of interactions and of dimensionality. A similar avalanche exponent is also observed in an independent work \cite{ruscher2019residual}, in which  the avalanche size is characterized using stress drops.
This exponent is clearly distinct from the one obtained at large plastic deformation in the stationary state using AQS or overdamped dynamics at zero temperature. In the latter case,  the avalanche exponent is larger than unity (close to 1.3 in simulations, to 1.5 according to depinning mean field predictions)  as  confirmed both by simulation and theoretical work \cite{PhysRevE.88.062206,PhysRevE.84.016115,dahmen2011simple}. 
This distinction is qualitatively consistent with experiments performed in metallic glasses \cite{krisponeit2014crossover} or simulations of  athermal cycling shear \cite{leishangthem2017yielding}, which shows a sharp transition from a transient regime to a steady state, and 
a different avalanche exponent  in these two states.
The large  avalanches in the steady state are system spanning and history-independent, they can be described as metabasin to metabasin transition on the potential energy landscape (PEL). In contrast, 
the small strains applied here perturb  the system  within a metabasin state of the PEL, and the avalanche is caused by basin to basin transition. Statistically, these two kinds of transition belong to two different universality classes.

As mentioned in the introduction, the Spigler Franz result is related to the hierarchical structure of the free energy landscape in a high dimension system. While the analysis is consistent with {the Franz-Spigler result} that the local minima within this landscape is marginally stable, it does not give any insight on the spatial structure and system size dependence of the associated events. In the following, we pursue the analysis of our simulation data in order to get some insight into this aspect, in an attempt to relate it to the view of marginal stability proposed by Lin and Wyart in the context of elasto-plastic models \cite{PhysRevX.6.011005}.

\begin{figure}[!htbp]
  \centering
  \includegraphics[width=1.0\columnwidth]{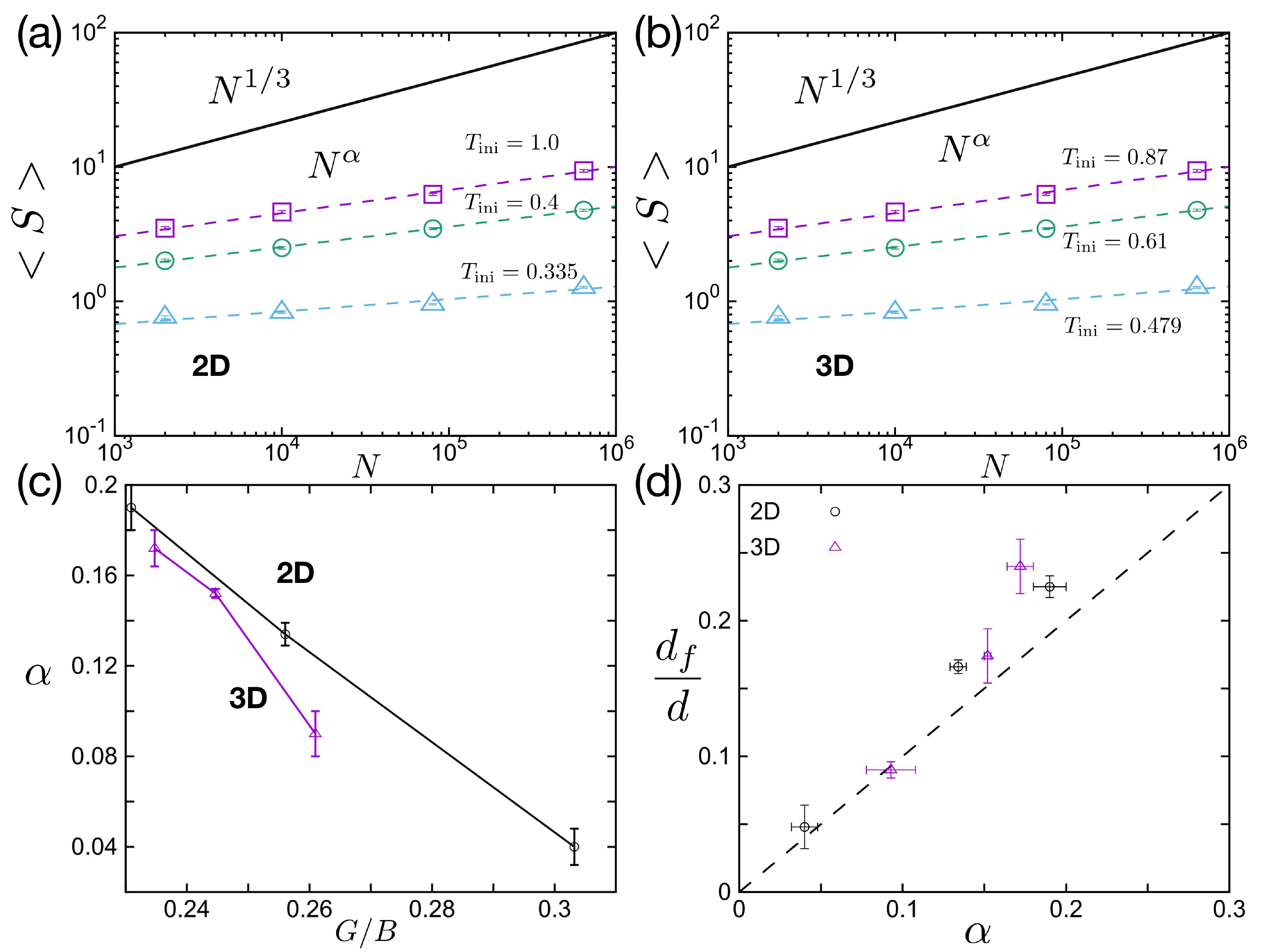}
  \caption{\textbf{Mean value of the avalanche size $<S>$} (a),(b) $<S>$ versus system size $N$ for different thermal  histories, in 2D and 3D systems. The dashed line is a fit by the equation $<S> \sim N^{\alpha}$ (c) finite size exponent $\alpha$ versus shear to bulk modulus ratio $G/B$. (d) Correlation between mean value exponent $\alpha$ and cutoff value exponent $d_f/d$ for various thermal histories and dimension.}
  \label{fig:4}
\end{figure}

\subsection*{Avalanche  mean size  $<S>$ and distribution cutoff $S_c$}

In addition to the avalanche exponent ($\tau$), the scaling parameters obtained by fitting the data plotted in figure \ref{fig:2} also provide  relevant information on avalanche statistics. The  fractal dimension $d_f$ characterizes the geometry of the avalanche event. We find that it  decreases when the initial stability of the system, characterized by $T_\text{ini}$, increases. 
Figure \ref{fig:4} shows that the mean value of avalanche size $<S>$ is also sensitive to system size and thermal history,  with a scaling exponent $\alpha$  larger than zero and  sub-extensive, as in the  steady state \cite{PhysRevLett.93.016001}, 
This result contrasts  the view that plastic activity  in the elastic regime of amorphous solids is localized and independent of system size \cite{PhysRevE.82.055103,PhysRevE.96.033002}.  Again, a check of the consistency of the results can be obtained by relating the values of the different exponents. As first discussed  by Lin et al \cite{lin2014scaling,PhysRevLett.115.168001}, for $1<\tau<2$, a scaling relation $\alpha =\frac{d_f}{d}(2-\tau)$ holds.
In our case, $\tau \approx 1$ , the scaling relation reduces  to $\alpha=d_f/d$, so that  $S_c \sim <S>$. Figure \ref{fig:4}(d) confirms the relation $\alpha=d_f/d$, and indeed the data can be collapsed equally well using  $S_c$ or $<S>$, as shown in Fig. S6.

Figure \ref{fig:4}(c) shows that the scaling exponent $\alpha$ is monotonically decreasing with the ratio of shear to bulk modulus $G/B$, which usually characterizes the ductility in amorphous materials \cite{lewandowski2005intrinsic,kumar2013critical,PhysRevE.95.022611}. This suggests a relation between ductility and  avalanche behavior in the elastic regime. We also note that, when  the  stability of the initial state  increases ($T_\text{ini}$ decreases), $\alpha$ becomes smaller and could eventually vanish.  
That situation suggests a transition from sub-extensive to localized avalanches, which could be connected with ductile to brittle transition dominated by initial stability\cite{Ozawa2018random}.
\begin{figure}[!htbp]
  \centering
  \includegraphics[width=1.0\columnwidth]{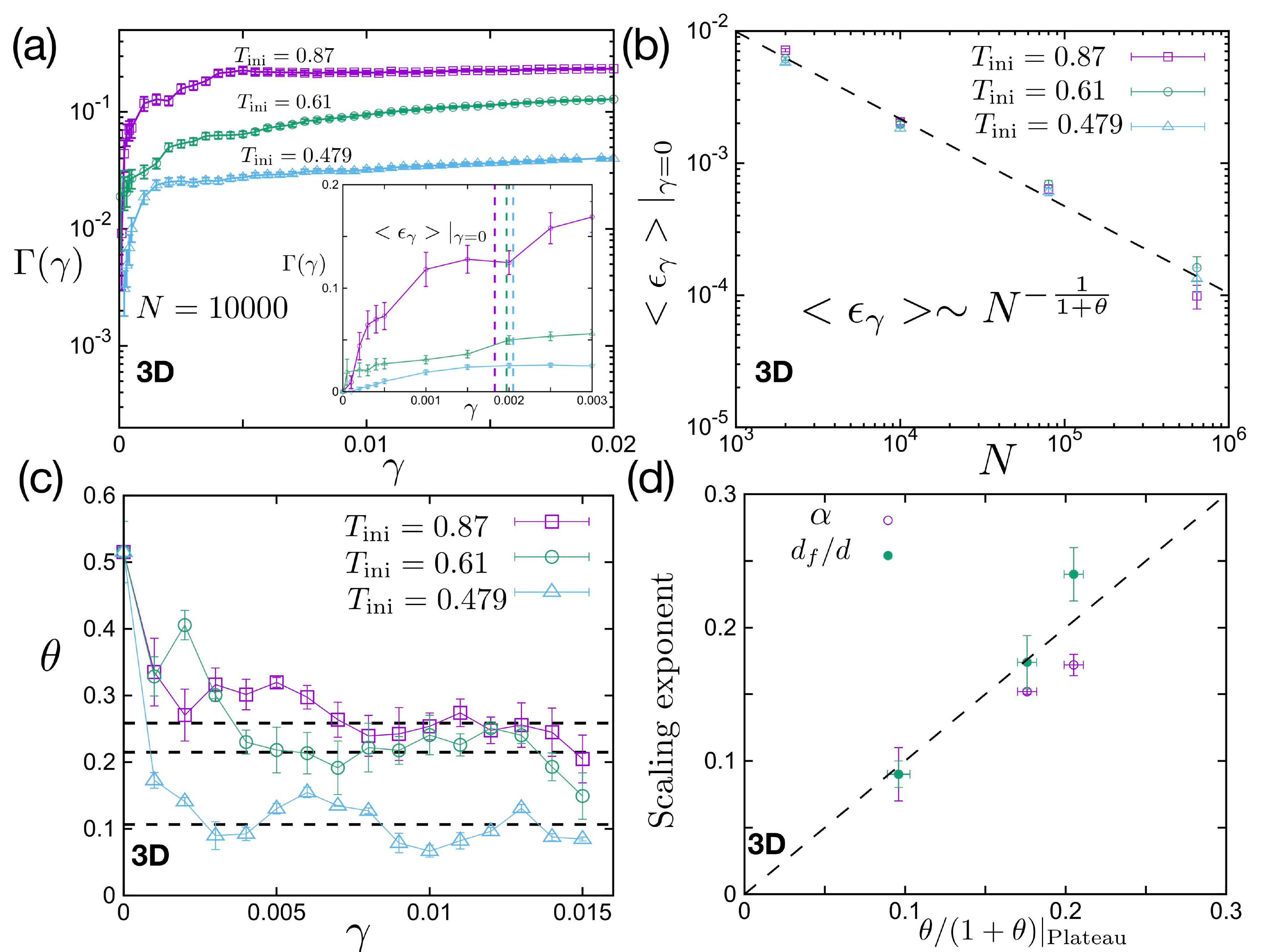}
  \caption{\textbf{The mean value of first avalanche event strain} (a)$\Gamma(\gamma)$ versus $\gamma$ for $N=10000$ in 3D system. Inset of (a): zooming in   $\Gamma(\gamma)$ vs $\gamma$ curve for different thermal histories, the vertical line shows the mean value of first avalanche event strain $<\epsilon_{\gamma}>|_{\gamma=0}$ , from left to right is $T_\text{ini}=0.87, 0.61, 0.479$, respectively. (b) $<\epsilon_{\gamma}>|_{\gamma=0}$ versus system size $N$ for different thermal histories in the 3D system, the dashed line is $N^{-0.66 \pm 0.02}$ where the exponent is fitted to the data. (c) Evolution of $\theta$ with  the strain window $\gamma$   for different thermal histories. The dashed line is the mean value of $\theta$ over the interval  $\gamma \in [0.005,0.015]$, we define this value as plateau value, from top to bottom,  $\theta_\text{Plateau}=0.26\pm 0.01,0.21\pm 0.01, 0.11 \pm 0.01$ for $T_\text{ini}=0.87, 0.61, 0.479$, respectively. (e) Correlation between finite size scaling exponents $\alpha$, $d_f/d$ and the exponent $\theta/(1+\theta)|_\text{Plateau}$ in the 3D system.}
  \label{fig:5}
\end{figure}

\subsection*{First avalanche event and pseudo-gap exponent $\theta$}
We now discuss the interpretation of our data within the context of elasto-plastic model, concentrating on the statistics of the first avalanche event observed upon applying strain to the system. To this end, we consider the evolution of the dissipated energy $\Gamma$ 
with the  $\gamma$ (figure \ref{fig:5}). For a given thermal history, $\Gamma$ as a function of strain varies slowly   except in the vicinity of the very first plastic event,  $\epsilon_{\gamma}|_{\gamma=0}$ \cite{PhysRevE.82.055103,lin2014density}. Here  $\epsilon_{\gamma}|_{\gamma}$ is defined as the incremental strain $\epsilon_{\gamma}$ needed to reach the next plastic event  after the system has been strained over an interval ${[0,\gamma]}$  
(see Fig. \ref{fig:f}(a)).
The behaviour shown in Figure \ref{fig:5}(a) interpolates between perfect elastic behavior without dissipation for $\gamma < \epsilon_{\gamma}|_{\gamma=0}$ and the regime $\epsilon_{\gamma}|_{\gamma=0} \ll \gamma \ll \gamma_{Y}$ in which dissipation is extensive and $\Gamma(\gamma)=\xi_1^2 \xi_2$. The strain scale for this crossover, $\epsilon_{\gamma}|_{\gamma=0}$, is a crucial quantity in the analysis of systems presenting marginal stability, as would be consistent with our observations for $\tau$. Indeed, in  a system presenting a pseudogap for low lying excitations of the form $P(x)\sim x^\theta$ (here $x$ is the strain associated with the excitation), extreme value statistics implies that the 
value of $\epsilon_{\gamma}|_{\gamma=0}$ scales with system size as $<\epsilon_{\gamma}> \sim N^{-\frac{1}{1+\theta}}$ \cite{Wyart2015marginal}. Here, a scaling $<\epsilon_{\gamma}>|_{\gamma=0} \sim N^{-0.66}$ is obtained as shown in figure \ref{fig:5}(b), implying $\theta\approx 1/2$. The latter value is consistent  with the  theoretical prediction of a  model of elastically interacting events \cite{PhysRevX.6.011005} and other simulation results \cite{PhysRevE.82.055103,PhysRevE.92.062302,lin2014density,barbot2018local,PhysRevE.98.063001}.

Within the framework of elasto-plastic models,  \cite{PhysRevLett.115.168001} the exponent $\theta$ can be related to  the exponent $\alpha$ that governs the dependence of the mean avalanche amplitude $<S>$ on system size. In the transient regime, the argument implies comparing  the number of avalanches over a stress interval, $M \sim \Delta \sigma / <\epsilon_{\gamma}> $, and the corresponding change in plastic strain, $<\Delta \gamma>_p  \sim M <S> /N $. As $\Delta \sigma/\Delta\gamma_p$ does not depend on system size 
(see Fig. \ref{fig:h} in SI Appendix), one obtains the relation between exponents $\alpha={\theta}/{(1+\theta)}$.

One is then faced with the paradoxical result, that the data indicates a significant dependence of $\alpha$ on the system preparation, while $\theta$ appears to have the universal value $1/2$. This discrepancy can be resolved by considering the fact that the avalanches that contribute to the definition of $\alpha$ are actually collected over a finite strain range, $\gamma \in [0,0.02]$. On the other hand, the value $\theta$ discussed previously involves only the very first event at $\gamma=0$,  $\epsilon_{\gamma}|_{\gamma=0}$. If we now extend the analysis to finite values of $\gamma$ and define a $\gamma$ dependent value of $\theta$ (characterizing the statistics of  $\epsilon_{\gamma}|_{\gamma}$), a very different value of $\theta$ is obtained, as illustrated in figure \ref{fig:5}(c). In fact, $\theta$ drops immediately from its initial value close to $1/2$ to a much lower value that depends on thermal history, and remains roughly constant over the whole strain interval. This behaviour corresponds to the one predicted by the  model of ref. \cite{PhysRevX.6.011005}, and the corresponding value of $\theta_\mathrm{Plateau}$ is perfectly correlated to the one obtained for $\alpha$, as illustrated in figure \ref{fig:5}(d). 

\section*{Discussion and Conclusions}

We have presented a detailed study of the avalanches that take place in the elastic portion of the stress-strain curve of an amorphous solid, or ``elastic avalanches''. We find several evidences that these avalanches have the characteristics expected for marginal states of dense amorphous packings, in particular, the avalanche exponent $\tau$ takes the value $\tau=1$ predicted by mean field theory for such packings \cite{PhysRevE.95.022139}. We propose to take this observation as   a possible  indication of a locally hierarchical energy landscape, 
and  explore the possibility that the corresponding events can be described within the framework of an elasto-plastic description in which marginal stability is associated with a pseudo gap in the distribution of excitations. Within this framework,  the exponent characterizing the pseudo-gap has a nontrivial evolution with strain, starting from a universal value 1/2 at zero strain and evolving rapidly towards a plateau that depends on thermal history. This behaviour also   corresponds to the  expectations of the model described in ref.  \cite{PhysRevX.6.011005}.

In addition, we find that the parameters characterizing avalanche distribution, energy dissipation and pseudo-gap are related by  three universal scaling relations $b+ 2 {d_f}/{d}=1$, $d_f/d=\alpha$, $\alpha=\theta/(1+\theta)$ and an identity $\Gamma=\xi_1^2 \xi_2$, regardless of dimension. While these results are established using quasi static simulations, the corresponding  analysis based on scaling arguments and energy conservation  should still hold at  finite strain rate and inertia.

In recent years, many efforts have been devoted to the identification of marginal stability in amorphous packings, and the present consensus \cite{Scalliet2019,PhysRevLett.122.255502} seems to be that this feature is observable only in systems with finite range, contact interactions, at relatively low packing fractions in the vicinity of the jamming point. It is therefore surprising that features of marginal stability are observed in systems with long range interactions and at packing fractions that are characteristic of high density glassy systems such as metallic glasses. We now tentatively explain this observation based on two observations. Firstly, the existence of a true, dissipation free elastic regime without avalanches depends on the manner in which the thermodynamic limit and the limit of zero strain are taken. The dissipation $\Gamma(\gamma)$ vanishes at small strain over a scale $\epsilon_\gamma|_{\gamma=0}$ that scales inversely to the system size, so that one has the two equalities: 
\begin{equation}
  \begin{cases}
	\lim \limits_{N \to \infty} \lim \limits_{\gamma \to 0} \Gamma(\gamma) &=0 \\
	\lim \limits_{\gamma \to 0} \lim \limits_{N \to \infty} \Gamma(\gamma) &=\xi^2_{1}\xi_2
  \end{cases}
  \label{eqn:n}
\end{equation}
The amorphous solid in the thermodynamic limit is therefore  intrinsically dissipative, as noted in previous  theoretical \cite{biroli2016breakdown} and simulation \cite{PhysRevE.83.061101} works. On the other hand, any finite system will have a finite range of ideal elastic behavior. Our study, however, indicates that this range will be crucially dependent on the thermal history and sample preparation. Indeed the behaviour observed for the pseudogap exponent and schematically summarized in 
Fig. \ref{fig:g},
as well as the behaviour of the exponent $\alpha$ (figure \ref{fig:3}),  indicates that  as the system is better annealed the range of elastic behavior will rapidly increase and the avalanches will become more fractal, with $\alpha = d_f/d$ approaching zero. As a result, it can be expected that in very well annealed systems such as those studied in refs \cite{Scalliet2019,PhysRevLett.122.255502,PhysRevE.98.063001} the size needed for observing large scale avalanches at small strains could be prohibitively large, so that observed excitations are limited to localised defects. The need to use larger sizes to properly describe the scaling behavior in the response of highly annealed systems was also pointed out in ref. \cite{PhysRevE.98.063001}. Whether or not there is an actual transition where $\alpha$ and {$\theta_\text{Plateau}$} vanish as a function of initial annealing conditions is an issue that cannot be addressed here, although this may be consistent with the idea of a sharp change from ductile to brittle behavior described in ref. \cite{Ozawa2018random}.
\subsection*{Methods}
\label{sec:Methods}
\subsubsection*{Sample preparation}
we use two well-studied glass-forming models to investigate the avalanche behavior within elastic regime: one is  2D Lenard-Jones binary model\cite{barbot2018local}, and the other is 3D Lenard-Jones binary model\cite{PhysRevLett.73.1376} with force shift\cite{toxvaerd2011communication}. All the units are reduced by the mass $m$, length scale $\sigma$, and energy $\epsilon$. The number density is fixed at $\rho=1.02$ and $1.20$ for 2D and 3D systems, respectively.
The number ratio between large ($N_{L}$) and small ($N_{S}$)  atoms is $N_{L}:N_{S}=(1+\sqrt{5})/4$ in 2D and $80:20$ in 3D.
We first annealed the sample to equilibrium state at $T_{\text{ini}}=0.335,0.4,1.0$ in 2D 
(see Fig. \ref{fig:0} in SI Appendix), $T_{\text{ini}}=0.479,0.61,0.87$ in 3D  using the  NVT ensemble, respectively. The temperature was controlled by a Nos\'e-Hoover thermostat \cite{nose1984unified} with periodic boundary conditions. The energy was then  minimized  to obtain the inherent structure at zero temperature, and we use $T_{\text{ini}}$ to represent the thermal history of each system.
In our two different LJ system,
we can take as a reference the  mode-coupling temperature $T_\text{MCT}$,  where
$T_\text{MCT}=0.325$ in 2D \cite{barbot2018local} and ${T_\text{MCT}=0.435}$ in 3D \cite{PhysRevE.52.4134}. All the simulations were conducted with the molecular dynamics simulation software: LAMMPS\cite{plimpton1995fast}.
\subsubsection*{Avalanche statistics in the elastic regime}
To investigate  finite size effects and the statistics of  avalanche distribution, we prepared series of  samples with different sizes  for each $T_\text{ini}$: in 2D , we used 2000 independent samples for $N=200$, 500 samples for $N=1000$, 100 samples for $N=10000$, 50 samples for $N=100000$ and in 3D:  50 samples for $N=2000$,20 samples for $N=10000$, 10 samples for $N=80000$, 1 sample for $N=640000$. 
Due to the  complexity of potential energy landscape, the avalanche events  
are highly depended on the deformation direction\cite{gendelman2015shear}, 
see Fig. \ref{fig:b} in SI Appendix,
we used the directional simple shear protocol to improve the statistics, in which a simple shear deformation gradient was used in different direction. As illustrated  in 
Fig. \ref{fig:a}, in the 2D system, 12 directions from $0$ to $\pi$ in xy plane are used for each sample, in the 3D system, simple shear deformation were applied in the xy ,xz, yz planes, respectively, and with again 12 directions from $0$ to $\pi$ in each plane.

Although strain and stress are tensors, for the simple shear deformation, the shear strain and shear stress dominate the mechanical deformation,  then we describe the deformation using the scalars $\gamma$ and $\tau_{\theta}$. 
As shown in Fig. \ref{fig:b},
we used an athermal quasistatic shear protocol to deform the sample. First the sample is affinely sheared by a small step strain, then the sample was minimized at deformed strain, repeated the process until the total strain reaches the desired value. The step strain both in 2D and 3D is $ \Delta \gamma =10^{-5}$ for all systems except the largest sample in 3D, where we used $\Delta \gamma = 2\times 10^{-6}$.
During the avalanche event, there is a stress drop and energy drop 
and  we define the avalanche size $S$ as
\begin{equation}
  S = N(\Delta U+\Delta \gamma \tau_{\theta}/\rho) 
\end{equation}
where $\Delta U$ is the potential energy drop per atom during avalanche, $\Delta \gamma$ is the strain step, $\tau_{\theta}$ is the  stress just before the avalanche, $\rho$ is the number density.
We use $S>0.01$ as a threshold to recognize  avalanche events. We have tested different thresholds from 0.01 to 0.1, with qualitatively similar results.
Following ref. \cite{PhysRevE.88.062206}, we define the avalanche number at a given avalanche size and system size per unit strain as $R(S,N,T_\text{ini})$. 
Note that both the avalanche number and avalanche size in the elastic regime not only depend on the system size $N$, but also  depend on the thermal history $T_\text{ini}$
(see Fig. \ref{fig:c} in SI Appendix)
\showmatmethods{}
\subsection*{Data deposition}
All data relevant to this paper are available at \url{https://doi.org/10.17605/OSF.IO/U6PYF}.
\acknow{
We thank Silvio Franz, Stefano Spigler, Matthieu Wyart and Ludovic Berthier for useful discussions. This work is supported by (B.S.S and P.G.F) the NSF of China (Grant Nos.51601009,Nos.51571011,Nos.U1930402),the MOST 973 Program (No.2015CB856800). B.S.S and P.F.G acknowledges the computational support from the Beijing Computational Science Research Center (CSRC). J.L.B is supported by Institut Universitaire de France.}
\showacknow{} 

\begin{thebibliography}{44}
\providecommand{\natexlab}[1]{#1}
\providecommand{\url}[1]{\texttt{#1}}
\expandafter\ifx\csname urlstyle\endcsname\relax
  \providecommand{\doi}[1]{doi: #1}\else
  \providecommand{\doi}{doi: \begingroup \urlstyle{rm}\Url}\fi

\bibitem[Ozawa et~al.(2018)Ozawa, Berthier, Biroli, Rosso, and
  Tarjus]{Ozawa2018random}
Misaki Ozawa, Ludovic Berthier, Giulio Biroli, Alberto Rosso, and Gilles
  Tarjus.
\newblock Random critical point separates brittle and ductile yielding
  transitions in amorphous materials.
\newblock \emph{Proc. Natl. Acad. Sci. U.S.A.}, 115\penalty0 (26):\penalty0
  6656--6661, 2018.
\newblock ISSN 0027-8424.

\bibitem[Berthier et~al.(2016)Berthier, Charbonneau, Jin, Parisi, Seoane, and
  Zamponi]{berthier2016growing}
Ludovic Berthier, Patrick Charbonneau, Yuliang Jin, Giorgio Parisi, Beatriz
  Seoane, and Francesco Zamponi.
\newblock Growing timescales and lengthscales characterizing vibrations of
  amorphous solids.
\newblock \emph{Proceedings of the National Academy of Sciences}, 113\penalty0
  (30):\penalty0 8397--8401, 2016.

\bibitem[Charbonneau et~al.(2017)Charbonneau, Kurchan, Parisi, Urbani, and
  Zamponi]{Charbonneau2017}
Patrick Charbonneau, Jorge Kurchan, Giorgio Parisi, Pierfrancesco Urbani, and
  Francesco Zamponi.
\newblock Glass and jamming transitions: From exact results to
  finite-dimensional descriptions.
\newblock \emph{Annu. Rev. Condens. Matter Phys.}, 8:\penalty0 265--288, 2017.

\bibitem[Scalliet et~al.(2019)Scalliet, Berthier, and Zamponi]{Scalliet2019}
Camille Scalliet, Ludovic Berthier, and Francesco Zamponi.
\newblock Marginally stable phases in mean-field structural glasses.
\newblock \emph{Phys. Rev. E}, 99:\penalty0 012107, Jan 2019.
\newblock \doi{10.1103/PhysRevE.99.012107}.
\newblock URL \url{https://link.aps.org/doi/10.1103/PhysRevE.99.012107}.

\bibitem[Scalliet and Berthier(2019)]{PhysRevLett.122.255502}
Camille Scalliet and Ludovic Berthier.
\newblock Rejuvenation and memory effects in a structural glass.
\newblock \emph{Phys. Rev. Lett.}, 122:\penalty0 255502, 2019.

\bibitem[Liao and Berthier(2019)]{PhysRevX.9.011049}
Qinyi Liao and Ludovic Berthier.
\newblock Hierarchical landscape of hard disk glasses.
\newblock \emph{Phys. Rev. X}, 9:\penalty0 011049, Mar 2019.
\newblock \doi{10.1103/PhysRevX.9.011049}.
\newblock URL \url{https://link.aps.org/doi/10.1103/PhysRevX.9.011049}.

\bibitem[Franz and Spigler(2017)]{PhysRevE.95.022139}
S.~Franz and S.~Spigler.
\newblock Mean-field avalanches in jammed spheres.
\newblock \emph{Phys. Rev. E}, 95:\penalty0 022139, 2017.

\bibitem[Papakonstantopoulos et~al.(2008)Papakonstantopoulos, Riggleman,
  Barrat, and de~Pablo]{Papakon2008}
George~J. Papakonstantopoulos, Robert~A. Riggleman, Jean-Louis Barrat, and
  Juan~J. de~Pablo.
\newblock Molecular plasticity of polymeric glasses in the elastic regime.
\newblock \emph{Phys. Rev. E}, 77:\penalty0 041502, Apr 2008.
\newblock \doi{10.1103/PhysRevE.77.041502}.
\newblock URL \url{https://link.aps.org/doi/10.1103/PhysRevE.77.041502}.

\bibitem[Antonaglia et~al.(2014)Antonaglia, Wright, Gu, Byer, Hufnagel,
  LeBlanc, Uhl, and Dahmen]{PhysRevLett.112.155501}
James Antonaglia, Wendelin~J. Wright, Xiaojun Gu, Rachel~R. Byer, Todd~C.
  Hufnagel, Michael LeBlanc, Jonathan~T. Uhl, and Karin~A. Dahmen.
\newblock Bulk metallic glasses deform via slip avalanches.
\newblock \emph{Phys. Rev. Lett.}, 112:\penalty0 155501, 2014.

\bibitem[Denisov et~al.(2017)Denisov, L{\H{o}}rincz, Wright, Hufnagel, Nawano,
  Gu, Uhl, Dahmen, and Schall]{denisov2017universal}
Dmitry~V Denisov, Kinga~A L{\H{o}}rincz, Wendelin~J Wright, Todd~C Hufnagel,
  Aya Nawano, Xiaojun Gu, Jonathan~T Uhl, Karin~A Dahmen, and Peter Schall.
\newblock Universal slip dynamics in metallic glasses and granular
  matter--linking frictional weakening with inertial effects.
\newblock \emph{Sci. Rep.}, 7:\penalty0 43376, 2017.

\bibitem[Lagogianni et~al.(2018)Lagogianni, Liu, Martens, and
  Samwer]{lagogianni2018plastic}
Alexandra~E Lagogianni, Chen Liu, Kirsten Martens, and Konrad Samwer.
\newblock Plastic avalanches in the so-called elastic regime of metallic
  glasses.
\newblock \emph{Eur. Phys. J. B}, 91\penalty0 (6):\penalty0 104, 2018.

\bibitem[Jin et~al.(2018)Jin, Urbani, Zamponi, and Yoshino]{Jin2018stability}
Yuliang Jin, Pierfrancesco Urbani, Francesco Zamponi, and Hajime Yoshino.
\newblock A stability-reversibility map unifies elasticity, plasticity,
  yielding, and jamming in hard sphere glasses.
\newblock \emph{Sci. Adv.}, 4\penalty0 (12), 2018.

\bibitem[Maloney and Lema\^itre(2004)]{PhysRevLett.93.016001}
Craig Maloney and Ana\"el Lema\^itre.
\newblock Subextensive scaling in the athermal, quasistatic limit of amorphous
  matter in plastic shear flow.
\newblock \emph{Phys. Rev. Lett.}, 93:\penalty0 016001, 2004.

\bibitem[Karmakar et~al.(2010)Karmakar, Lerner, and
  Procaccia]{PhysRevE.82.055103}
Smarajit Karmakar, Edan Lerner, and Itamar Procaccia.
\newblock Statistical physics of the yielding transition in amorphous solids.
\newblock \emph{Phys. Rev. E}, 82:\penalty0 055103, 2010.

\bibitem[Fan et~al.(2017)Fan, Wang, Zhang, Liu, Schroers, Shattuck, and
  O'Hern]{PhysRevE.95.022611}
Meng Fan, Minglei Wang, Kai Zhang, Yanhui Liu, Jan Schroers, Mark~D. Shattuck,
  and Corey~S. O'Hern.
\newblock Effects of cooling rate on particle rearrangement statistics: Rapidly
  cooled glasses are more ductile and less reversible.
\newblock \emph{Phys. Rev. E}, 95:\penalty0 022611, 2017.

\bibitem[Krisponeit et~al.(2014)Krisponeit, Pitikaris, Avila, K{\"u}chemann,
  Kr{\"u}ger, and Samwer]{krisponeit2014crossover}
Jon-Olaf Krisponeit, Sebastian Pitikaris, Karina~E Avila, Stefan K{\"u}chemann,
  Antje Kr{\"u}ger, and Konrad Samwer.
\newblock Crossover from random three-dimensional avalanches to correlated nano
  shear bands in metallic glasses.
\newblock \emph{Nat. Commun.}, 5:\penalty0 3616, 2014.

\bibitem[Leishangthem et~al.(2017)Leishangthem, Parmar, and
  Sastry]{leishangthem2017yielding}
Premkumar Leishangthem, Anshul D.~S. Parmar, and Srikanth Sastry.
\newblock The yielding transition in amorphous solids under oscillatory shear
  deformation.
\newblock \emph{Nat. Commun.}, 8:\penalty0 14653, 2017.

\bibitem[Regev et~al.(2015)Regev, Weber, Reichhardt, Dahmen, and
  Lookman]{regev2015reversibility}
Ido Regev, John Weber, Charles Reichhardt, Karin~A Dahmen, and Turab Lookman.
\newblock Reversibility and criticality in amorphous solids.
\newblock \emph{Nat. Commun.}, 6:\penalty0 8805, 2015.

\bibitem[Peng et~al.(2019)Peng, Zhang, Yang, Li, Zhang, Liu, Yu, and
  Samwer]{peng2019anomalous}
Si-Xu Peng, Cheng Zhang, Chong Yang, Ran Li, Tao Zhang, Lin Liu, Hai-Bin Yu,
  and Konrad Samwer.
\newblock Anomalous nonlinear damping in metallic glasses: Signature of
  elasticity breakdown.
\newblock \emph{J. Chem. Phys.}, 150\penalty0 (11):\penalty0 111104, 2019.

\bibitem[Lin and Wyart(2016)]{PhysRevX.6.011005}
Jie Lin and Matthieu Wyart.
\newblock Mean-field description of plastic flow in amorphous solids.
\newblock \emph{Phys. Rev. X}, 6:\penalty0 011005, 2016.

\bibitem[Salerno et~al.(2012)Salerno, Maloney, and
  Robbins]{PhysRevLett.109.105703}
K.~Salerno, Craig Maloney, and Mark Robbins.
\newblock Avalanches in strained amorphous solids: Does inertia destroy
  critical behavior?
\newblock \emph{Phys. Rev. Lett.}, 109:\penalty0 105703, 2012.

\bibitem[Shi and Falk(2005)]{PhysRevLett.95.095502}
Yunfeng Shi and Michael~L. Falk.
\newblock Strain localization and percolation of stable structure in amorphous
  solids.
\newblock \emph{Phys. Rev. Lett.}, 95:\penalty0 095502, 2005.

\bibitem[Salerno and Robbins(2013)]{PhysRevE.88.062206}
K.~Michael Salerno and Mark~O. Robbins.
\newblock Effect of inertia on sheared disordered solids: Critical scaling of
  avalanches in two and three dimensions.
\newblock \emph{Phys. Rev. E}, 88:\penalty0 062206, 2013.

\bibitem[Ruscher and Rottler(2019)]{ruscher2019residual}
C{\'e}line Ruscher and J{\"o}rg Rottler.
\newblock Residual stress distributions in athermally deformed amorphous solids
  from atomistic simulations.
\newblock \emph{arXiv preprint arXiv:1908.01081}, 2019.

\bibitem[Talamali et~al.(2011)Talamali, Pet\"aj\"a, Vandembroucq, and
  Roux]{PhysRevE.84.016115}
Mehdi Talamali, Viljo Pet\"aj\"a, Damien Vandembroucq, and St\'ephane Roux.
\newblock Avalanches, precursors, and finite-size fluctuations in a mesoscopic
  model of amorphous plasticity.
\newblock \emph{Phys. Rev. E}, 84:\penalty0 016115, 2011.

\bibitem[Dahmen et~al.(2011)Dahmen, Ben-Zion, and Uhl]{dahmen2011simple}
Karin~A Dahmen, Yehuda Ben-Zion, and Jonathan~T Uhl.
\newblock A simple analytic theory for the statistics of avalanches in sheared
  granular materials.
\newblock \emph{Nat. Phys.}, 7\penalty0 (7):\penalty0 554--557, 2011.

\bibitem[Lin and Zheng(2017)]{PhysRevE.96.033002}
Jie Lin and Wen Zheng.
\newblock Universal scaling of the stress-strain curve in amorphous solids.
\newblock \emph{Phys. Rev. E}, 96:\penalty0 033002, 2017.

\bibitem[Lin et~al.(2014{\natexlab{a}})Lin, Lerner, Rosso, and
  Wyart]{lin2014scaling}
Jie Lin, Edan Lerner, Alberto Rosso, and Matthieu Wyart.
\newblock Scaling description of the yielding transition in soft amorphous
  solids at zero temperature.
\newblock \emph{Proc. Natl. Acad. Sci. U.S.A.}, 111\penalty0 (40):\penalty0
  14382--14387, 2014{\natexlab{a}}.

\bibitem[Lin et~al.(2015)Lin, Gueudr\'e, Rosso, and
  Wyart]{PhysRevLett.115.168001}
Jie Lin, Thomas Gueudr\'e, Alberto Rosso, and Matthieu Wyart.
\newblock Criticality in the approach to failure in amorphous solids.
\newblock \emph{Phys. Rev. Lett.}, 115:\penalty0 168001, 2015.

\bibitem[Lewandowski et~al.(2005)Lewandowski, Wang, and
  Greer]{lewandowski2005intrinsic}
JJ~Lewandowski, WH~Wang, and AL~Greer.
\newblock Intrinsic plasticity or brittleness of metallic glasses.
\newblock \emph{Philos. Mag. Lett.}, 85\penalty0 (2):\penalty0 77--87, 2005.

\bibitem[Kumar et~al.(2013)Kumar, Neibecker, Liu, and
  Schroers]{kumar2013critical}
Golden Kumar, Pascal Neibecker, Yan~Hui Liu, and Jan Schroers.
\newblock Critical fictive temperature for plasticity in metallic glasses.
\newblock \emph{Nat. Commun.}, 4:\penalty0 1536, 2013.

\bibitem[Lin et~al.(2014{\natexlab{b}})Lin, Saade, Lerner, Rosso, and
  Wyart]{lin2014density}
Jie Lin, Alaa Saade, Edan Lerner, Alberto Rosso, and Matthieu Wyart.
\newblock On the density of shear transformations in amorphous solids.
\newblock \emph{EPL}, 105\penalty0 (2):\penalty0 26003, 2014{\natexlab{b}}.

\bibitem[M\"uller and Wyart(2015)]{Wyart2015marginal}
Markus M\"uller and Matthieu Wyart.
\newblock Marginal stability in structural, spin, and electron glasses.
\newblock \emph{Annu. Rev. Condens. Matter Phys.}, 6\penalty0 (1):\penalty0
  177--200, 2015.

\bibitem[Hentschel et~al.(2015)Hentschel, Jaiswal, Procaccia, and
  Sastry]{PhysRevE.92.062302}
H.~G.~E. Hentschel, Prabhat~K. Jaiswal, Itamar Procaccia, and Srikanth Sastry.
\newblock Stochastic approach to plasticity and yield in amorphous solids.
\newblock \emph{Phys. Rev. E}, 92:\penalty0 062302, 2015.

\bibitem[Barbot et~al.(2018)Barbot, Lerbinger, Hernandez-Garcia,
  Garc{\'\i}a-Garc{\'\i}a, Falk, Vandembroucq, and Patinet]{barbot2018local}
Armand Barbot, Matthias Lerbinger, Anier Hernandez-Garcia, Reinaldo
  Garc{\'\i}a-Garc{\'\i}a, Michael~L Falk, Damien Vandembroucq, and Sylvain
  Patinet.
\newblock Local yield stress statistics in model amorphous solids.
\newblock \emph{Phys. Rev. E}, 97\penalty0 (3):\penalty0 033001, 2018.

\bibitem[Lerner et~al.(2018)Lerner, Procaccia, Rainone, and
  Singh]{PhysRevE.98.063001}
Edan Lerner, Itamar Procaccia, Corrado Rainone, and Murari Singh.
\newblock Protocol dependence of plasticity in ultrastable amorphous solids.
\newblock \emph{Phys. Rev. E}, 98:\penalty0 063001, 2018.

\bibitem[Biroli and Urbani(2016)]{biroli2016breakdown}
Giulio Biroli and Pierfrancesco Urbani.
\newblock Breakdown of elasticity in amorphous solids.
\newblock \emph{Nat. Phys.}, 12\penalty0 (12):\penalty0 1130, 2016.

\bibitem[Hentschel et~al.(2011)Hentschel, Karmakar, Lerner, and
  Procaccia]{PhysRevE.83.061101}
H.~G.~E. Hentschel, Smarajit Karmakar, Edan Lerner, and Itamar Procaccia.
\newblock Do athermal amorphous solids exist?
\newblock \emph{Phys. Rev. E}, 83:\penalty0 061101, 2011.

\bibitem[Kob and Andersen(1994)]{PhysRevLett.73.1376}
Walter Kob and Hans~C. Andersen.
\newblock Scaling behavior in the $\beta$-relaxation regime of a supercooled
  lennard-jones mixture.
\newblock \emph{Phys. Rev. Lett.}, 73:\penalty0 1376--1379, 1994.

\bibitem[Toxvaerd and Dyre(2011)]{toxvaerd2011communication}
S{\o}ren Toxvaerd and Jeppe~C Dyre.
\newblock Communication: Shifted forces in molecular dynamics.
\newblock \emph{J. Chem. Phys.}, 134:\penalty0 081102, 2011.

\bibitem[Nos{\'e}(1984)]{nose1984unified}
Shuichi Nos{\'e}.
\newblock A unified formulation of the constant temperature molecular dynamics
  methods.
\newblock \emph{J. Chem. Phys.}, 81:\penalty0 511, 1984.

\bibitem[Kob and Andersen(1995)]{PhysRevE.52.4134}
Walter Kob and Hans~C. Andersen.
\newblock Testing mode-coupling theory for a supercooled binary lennard-jones
  mixture. ii. intermediate scattering function and dynamic susceptibility.
\newblock \emph{Phys. Rev. E}, 52:\penalty0 4134--4153, 1995.

\bibitem[Plimpton(1995)]{plimpton1995fast}
Steve Plimpton.
\newblock Fast parallel algorithms for short-range molecular dynamics.
\newblock \emph{J. Comput. Phys}, 117\penalty0 (1):\penalty0 1--19, 1995.

\bibitem[Gendelman et~al.(2015)Gendelman, Jaiswal, Procaccia, Gupta, and
  Zylberg]{gendelman2015shear}
Oleg Gendelman, Prabhat~K Jaiswal, Itamar Procaccia, Bhaskar~Sen Gupta, and
  Jacques Zylberg.
\newblock Shear transformation zones: State determined or protocol dependent?
\newblock \emph{EPL}, 109\penalty0 (1):\penalty0 16002, 2015.

\end{thebibliography}

\clearpage
\pagebreak
\section*{Supplamentary Information}
\renewcommand{\thefigure}{S\arabic{figure}}
\renewcommand{\theequation}{S\arabic{equation}}
\renewcommand{\thesection}{S\arabic{section}}
\renewcommand{\thetable}{S\arabic{table}}
\setcounter{figure}{0}
\setcounter{equation}{0}
\setcounter{section}{0}

\subsection*{Fitting parameters}
We have summarized all the fitting parameters of scaling collapse in the main text as follows:
\begin{table}[!htbp]
  \centering
  \begin{tabular}{ccccc}
	\hline
	$T_\text{ini}$  & ${d_f}/{d}$ &  $\alpha$ & $\xi_1$ & $\xi_1^2 \xi_2$  \\
	\hline
	1.0 & 0.225(8) & 0.19(1) & 1.2(1) & 0.196(5) \\
	0.4 & 0.166(5) & 0.134(5) & 0.84(4) & 0.093(5) \\
	0.335 & 0.050(2) & 0.040(8) & 0.46(7) & 0.013(1) \\
	\hline
  \end{tabular}
  \caption{Fit parameters for the  2D system, after scaling collapse , the master curve is $\sim x^{-\tau}f(x/x_c)$, where avalanche exponent $\tau=0.98 \pm 0.01$.}
  \label{tab:2D}
\end{table}
\begin{table}[!htbp]
  \centering
  \begin{tabular}{ccccc}
	\hline
	$T_\text{ini}$  & ${d_f}/{d}$ &  $\alpha$ & $\xi_1$ & $\xi_1^2 \xi_2$  \\
	\hline
	0.87 & 0.24(2) &  0.172(8) & 2.8(6) & 0.23(1) \\
	0.61 & 0.17(2) &  0.152(2) & 2.6(7) & 0.131(4) \\
	0.479 & 0.090(6) &  0.09(1) & 1.5(1) & 0.040(2) \\
	\hline
  \end{tabular}
  \caption{Fit parameters for the  3D system,  after scaling collapse , the master curve is $\sim x^{-\tau}f(x/x_c)$, where avalanche exponent $\tau=1.01 \pm 0.01$.}
  \label{tab:3D}
\end{table}

\begin{figure}[!htbp]
  \centering
  \includegraphics[width=1.0\columnwidth]{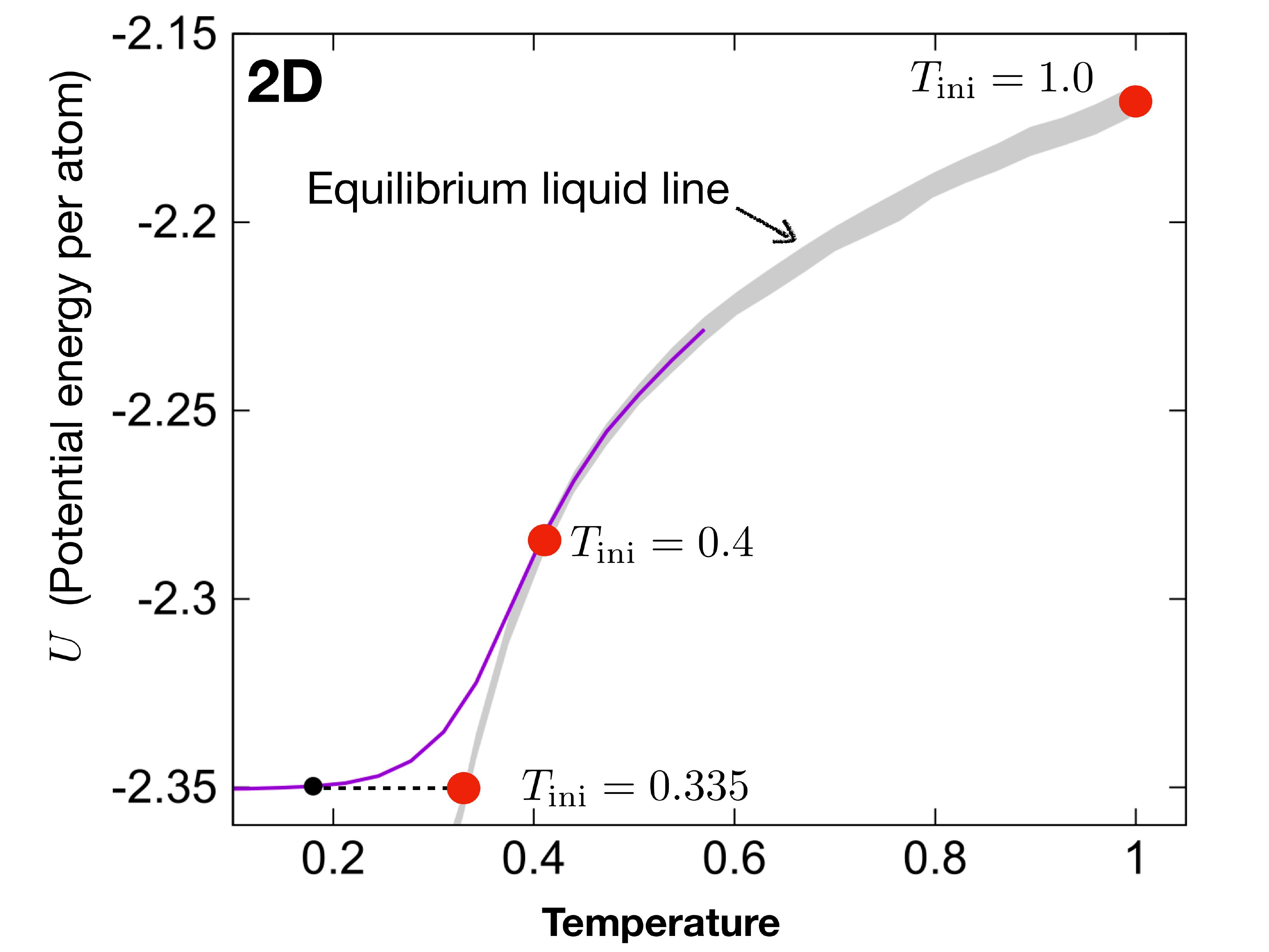}
  \caption{Thermal history of the 2D sample in the temperature and inherent structure potential energy diagram. The grey line is the equilibrium liquid line, and the red solid point is representative thermal history used in the article, the purple line is the liquid to glass transition line at a given  quench rate.}
  \label{fig:0}
\end{figure}
\begin{figure}[!htbp]
  \centering
  \includegraphics[width=1.0\columnwidth]{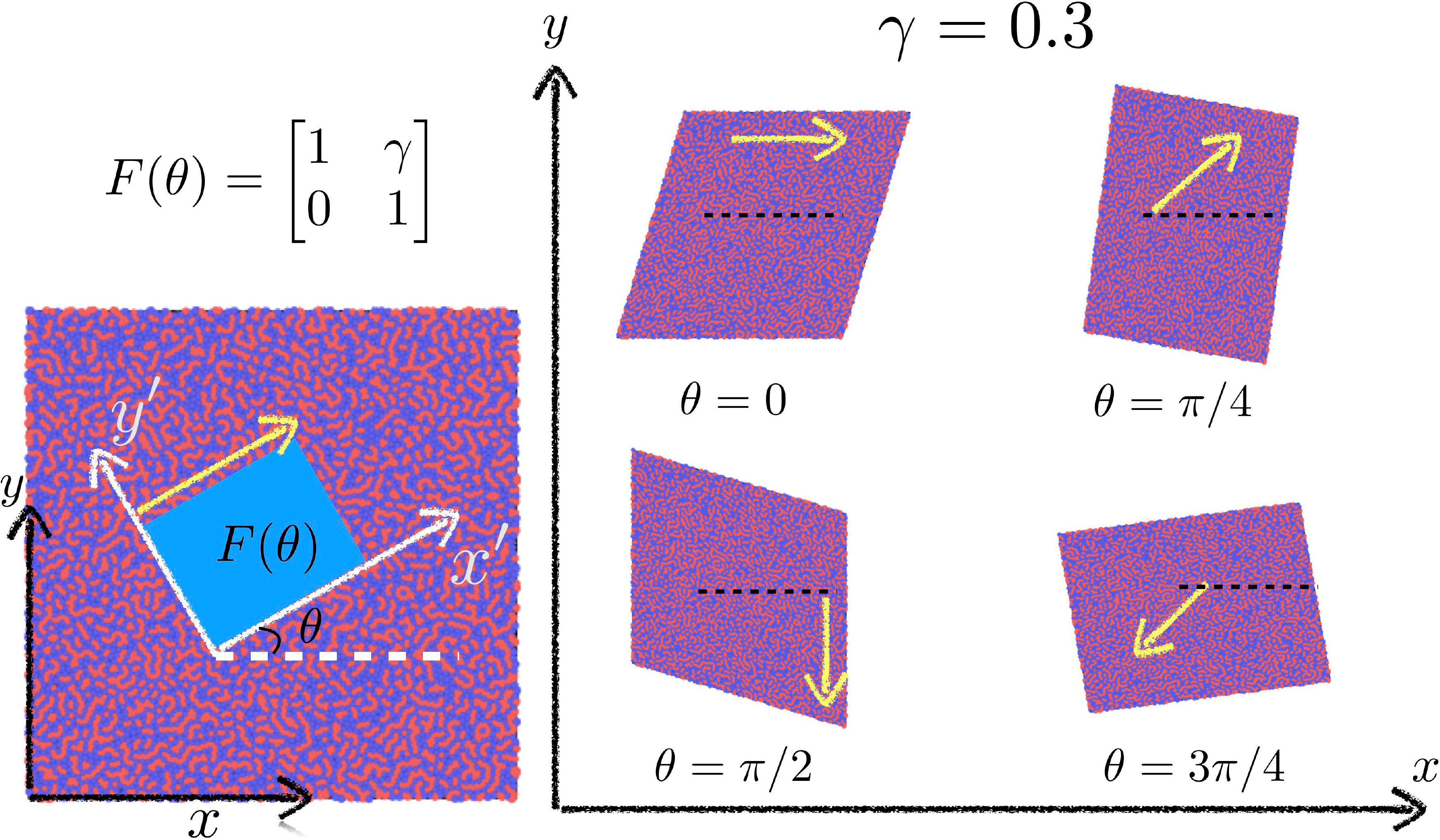}
  \caption{Schematic diagram of the deformations applied in athermal quasistatic shear,left panel: the simple shear deformation gradient in direction $\theta$, right panel: typical sample deformation  for four different directions at a given strain.} \label{fig:a}
\end{figure}

\begin{figure}[!htbp]
  \centering
  \includegraphics[width=1.0\columnwidth]{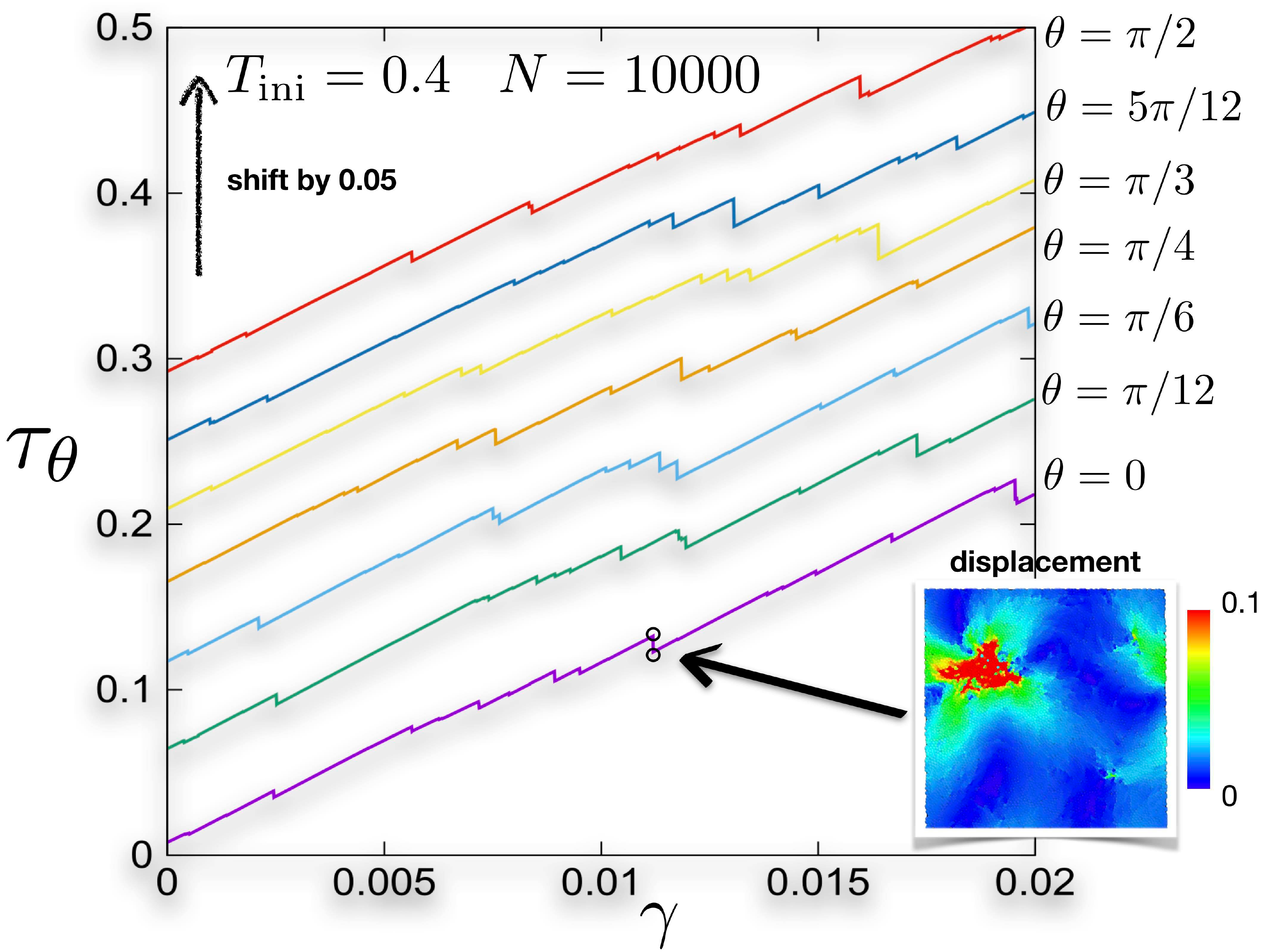}
  \caption{Strain stress curves for different loading directions for $N=10000$, $T_\text{ini}=0.4$ in the 2D system, each curve from bottom to up is shifted by 0.05. The inset shows a  color plot of the displacement field between two configurations separated by the stress drop displayed with  hollow circles in the strain stress curve.}
  \label{fig:b}
\end{figure}
\begin{figure}[!htbp]
  \centering
  \includegraphics[width=1.0\columnwidth]{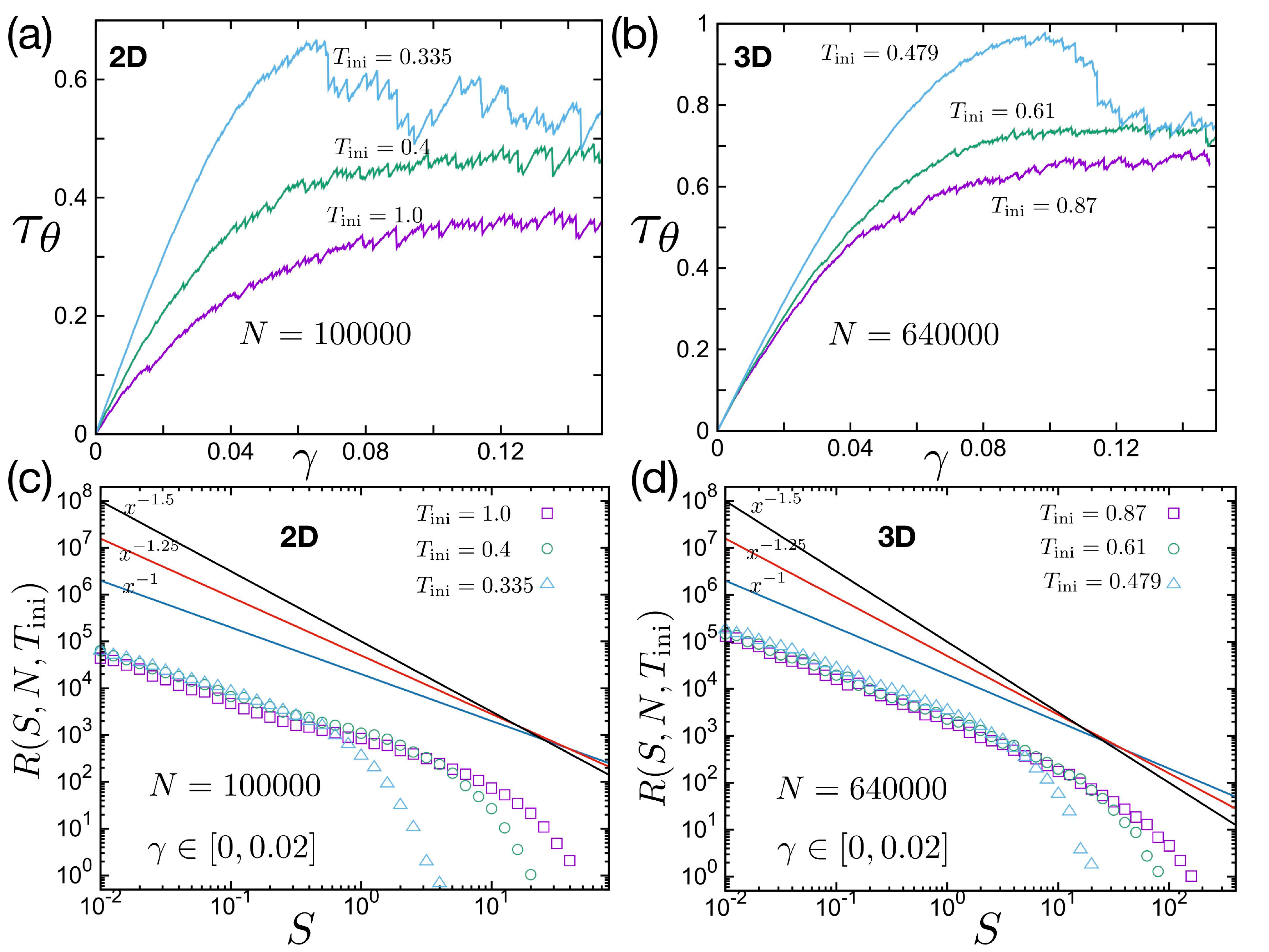}
  \caption{{Avalanche distribution in the  elastic regime for different thermal history:} (a),(b) typical stress strain curve of one sample for different thermal history in 2D and 3D system, respectively. (c),(d) The avalanche number distribution within $\gamma\in [0,0.02]$ for different thermal histories    in 2D and 3D. The solid lines show three possible avalanche exponents.}
  \label{fig:c}
\end{figure}

\begin{figure}[!htbp]
  \centering
  \includegraphics[width=1.0\columnwidth]{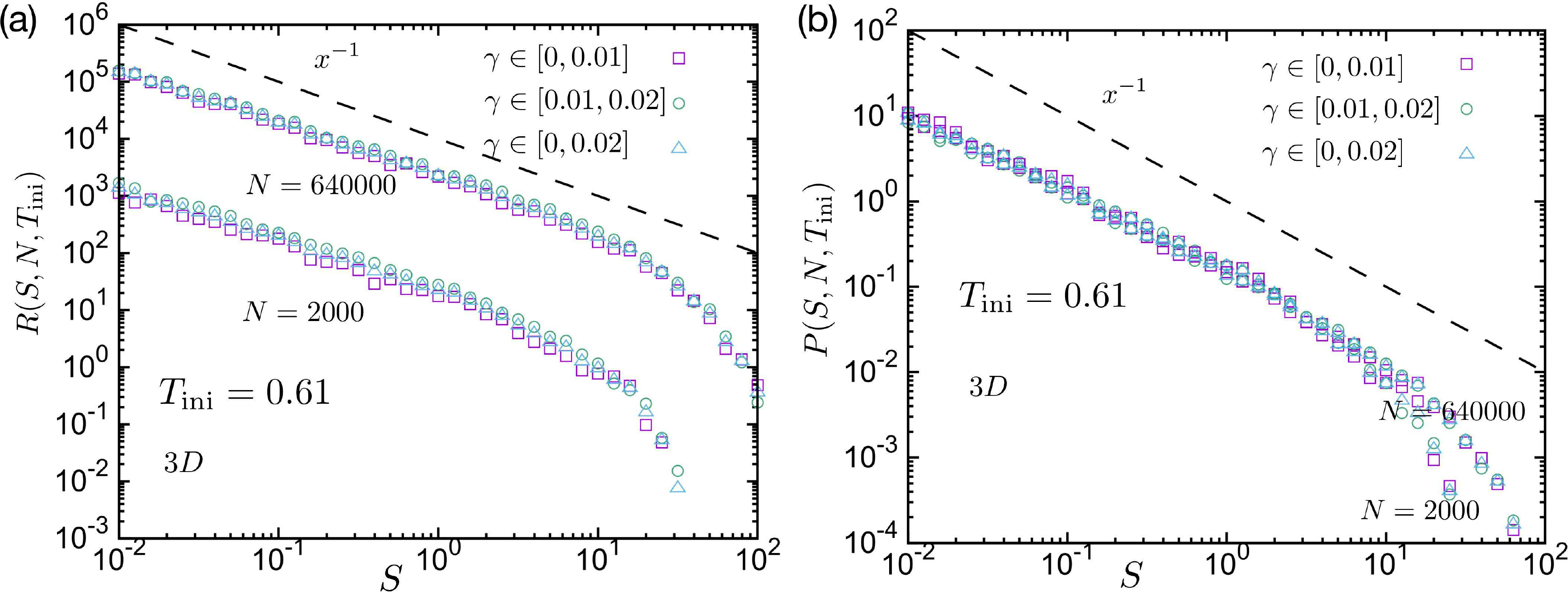}
  \caption{(a) Number of avalanches per unit strain  for  different strain intervals,   for the $N=2000$ and $N=640000$  3D system. (b) Probability distribution of avalanche sizes for the same parameters as in (a), the dashed line has a slope  $-1$. }
  \label{fig:d}
\end{figure}

\begin{figure}[!htbp]
  \centering
\includegraphics[width=1.0\columnwidth]{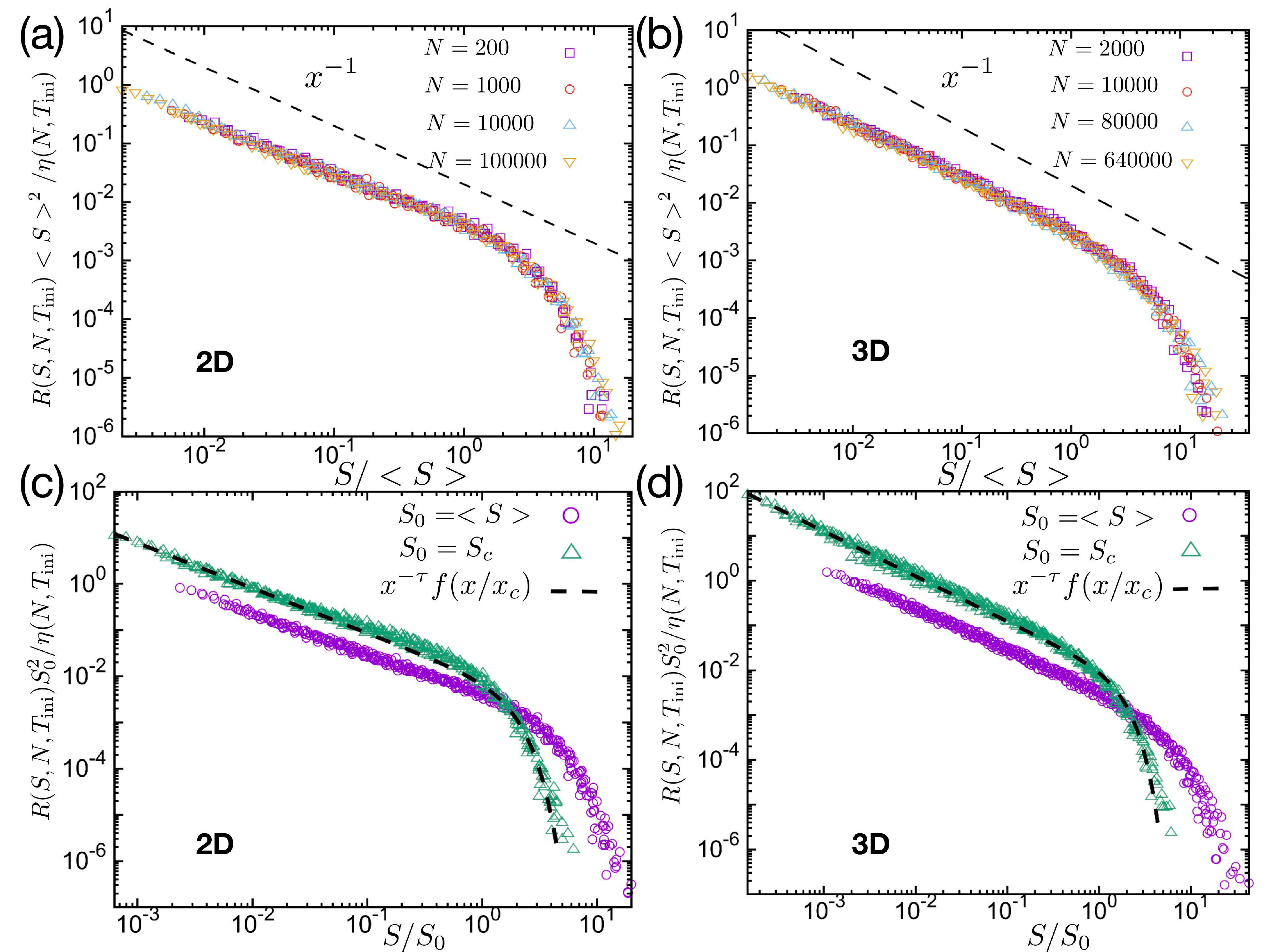}
  \caption{(a),(b) Scaling collapse of the avalanche distribution using the  average avalanche size in 2D and 3D systems, respectively. (c),(d) comparison between the  scaling collapse obtained using  average avalanche size $<S>$ or cutoff value $S_c$. The dashed line is a fit using the function  $Ax^{-\tau}f(x/x_c)$, where $f(x/x_c)=e^{-{x^2}/{x_c^2}}$, $\tau=0.98 \pm 0.01$,$1.01 \pm 0.01$ for 2D,3D systems, respectively.}
  \label{fig:e}
\end{figure}
\begin{figure}[!htbp]
  \centering
\includegraphics[width=1.0\columnwidth]{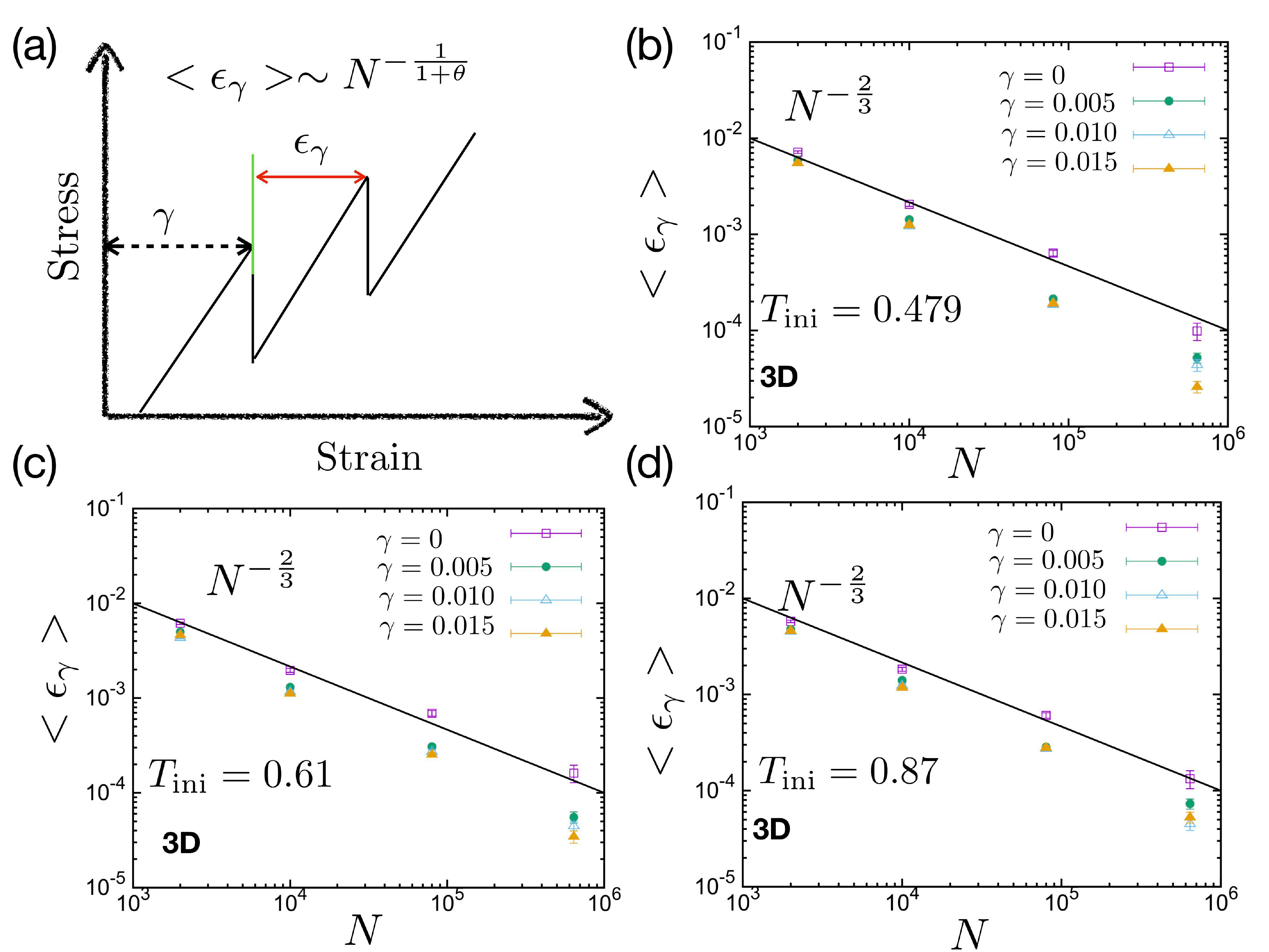}
  \caption{\textbf{Strain at the first plastic event following a previous strain, for   different previous strain intervals.} (a) Schematic representation of the strain at the first plastic event,  $\epsilon_\gamma$, for a given previous strain $\gamma$. (b),(c),(d) Mean value of $\epsilon_\gamma$ versus system size $N$ for different previous strain intervals  $\gamma$ for $T_\text{ini}=0.479, 0.61, 0.87$, respectively, the solid line is $N^{-\frac{2}{3}}$.}
  \label{fig:f}
\end{figure}
\begin{figure}[!htbp]
  \centering
  \includegraphics[width=1.0\columnwidth]{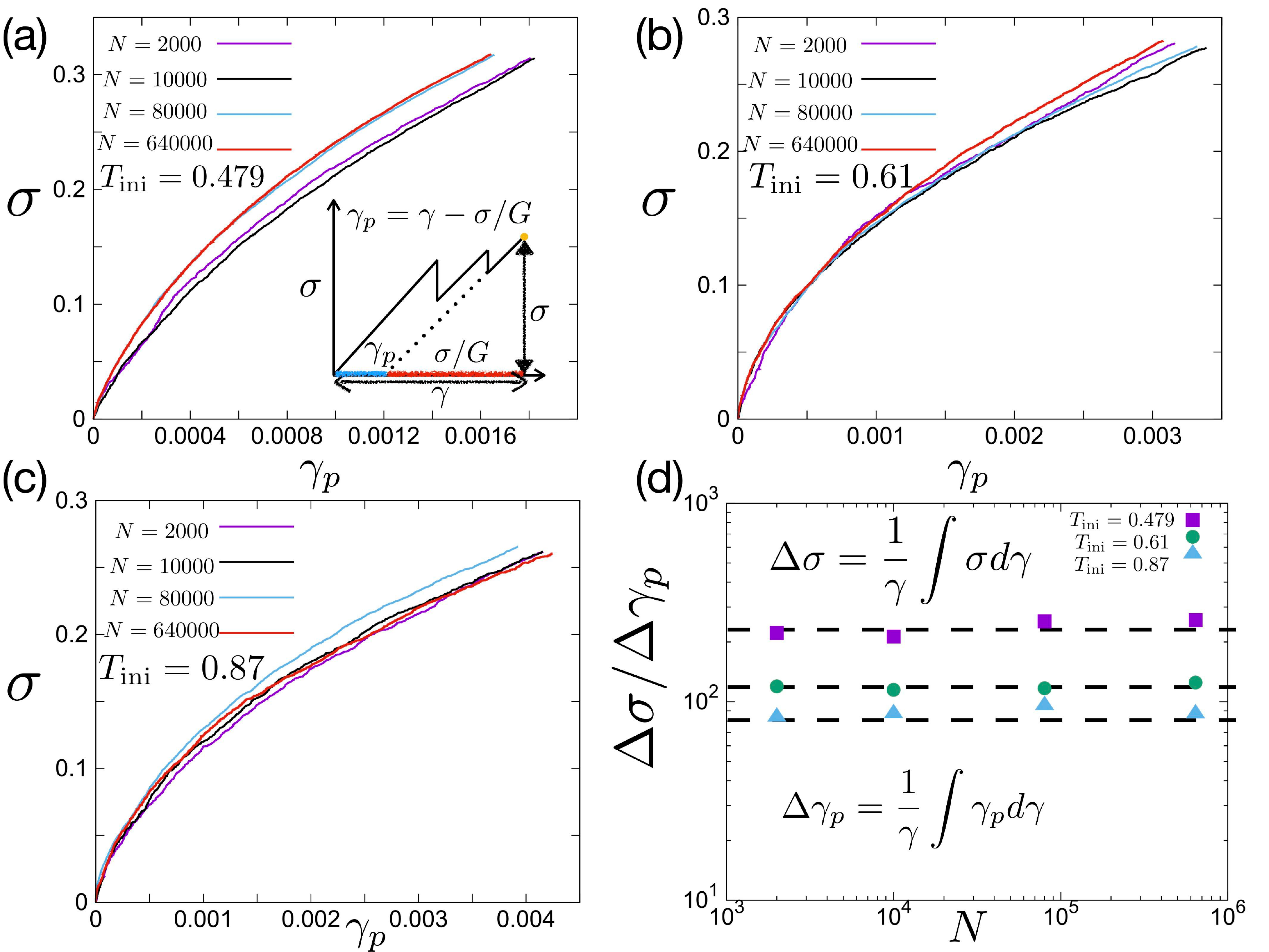}
  \caption{(a),(b),(c) plastic strain versus stress for different system sizes and thermal histories. The inset in panel (a)  shows how the plastic strain  $\gamma_{p}$ in obtained from the strain-stress curves, $\gamma_{p}=\gamma -\sigma/G$, where $\gamma$ is the total shear strain, $\sigma$ is the mean value of shear stress, and $G$ is the elastic shear modulus.  (d) the ratio of average stress and average plastic strain $\Delta \sigma/ \Delta \gamma_{p}$ versus system sizes for different thermal histories. The horizontal dashed line is the mean value of the ratio $\Delta \sigma/ \Delta \gamma_{p}$  for each thermal history. The strain interval is $\gamma \in [0,0.02]$.  }
  \label{fig:h}
\end{figure}
\begin{figure}[!htbp]
  \centering
  \includegraphics[width=1.0\columnwidth]{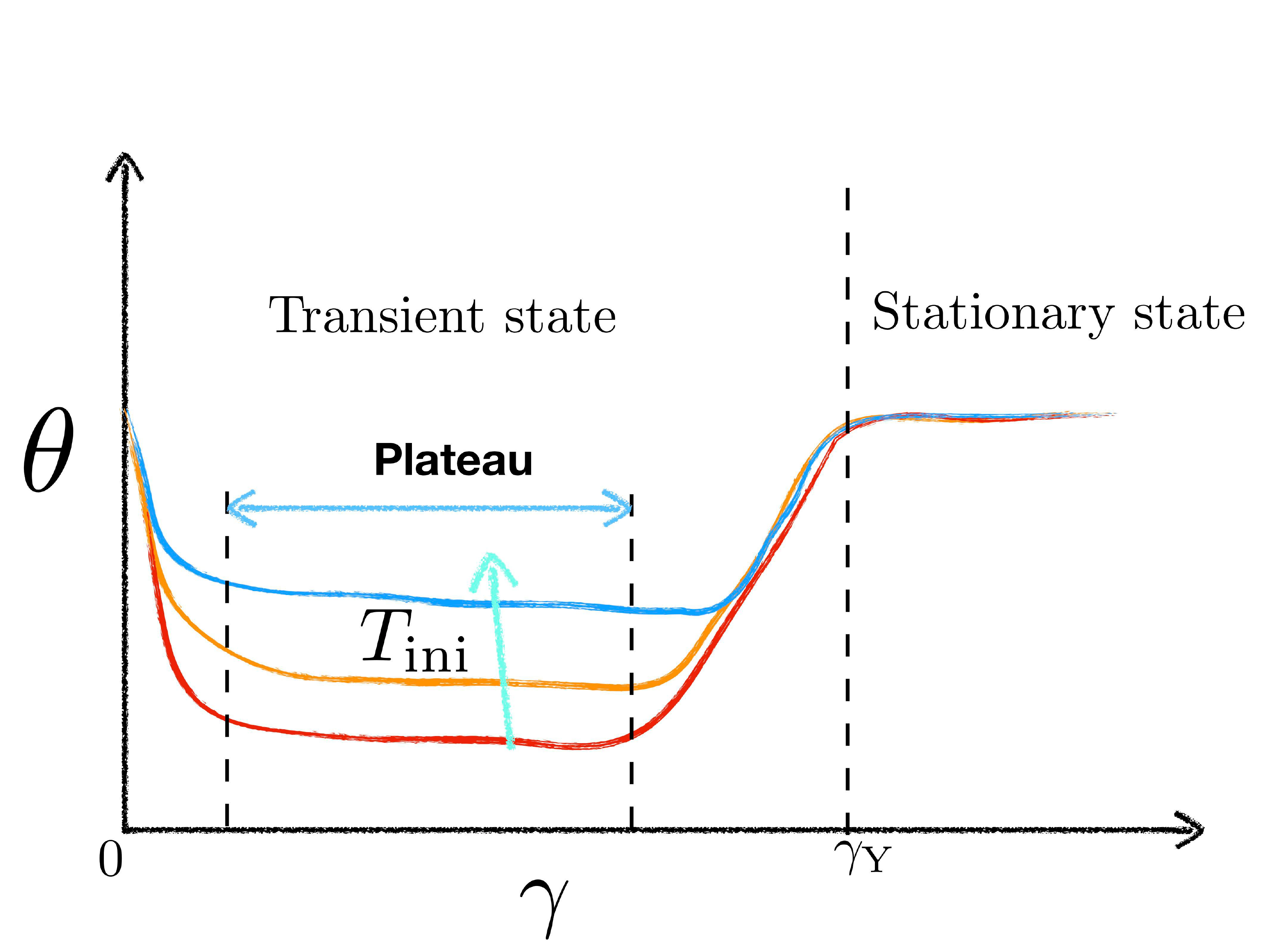}
  \caption{Schematic representation  of the non monotonic evolution of $\theta$ with strain $\gamma$ in the transient state and stationary state for different thermal histories.}
  \label{fig:g}
\end{figure}
\end{document}